\documentclass{article}
\usepackage[english]{babel}
\usepackage{amsmath, amssymb}
\usepackage{graphicx}
\usepackage{float}
\usepackage{array}
\usepackage{geometry}
\title{Inspection of ratcheting models for pathological error sensitivity and overparametrization}

\author{A. A. Kaygorodtseva, A. V. Shutov  \\
	Lavrentyev Institute of Hydrodynamics \\ Pr. Lavrentyeva 15, Novosibirsk 630090, Russia  \\
	Novosibirsk State University \\ Ul. Pirogova 1, Novosibirsk 630090, Russia \\
	}

\begin{document}
\maketitle

\begin{abstract}
Accurate analysis of plastic strain accumulation under stress-controlled cyclic loading is vital for numerous engineering applications. Typically, models of plastic ratcheting are calibrated against available experimental data. Since actual experiments are not exactly accurate, one should check the identification protocols for pathological dependencies on experimental errors. In this paper, a step-by-step algorithm is presented to estimate the sensitivities of identified material parameters.
As a part of the sensitivity analysis method, a new mechanics-based metric in the space of material parameters is proposed especially for ratcheting-related applications. The sensitivity of material parameters to experimental errors is estimated, based on this metric.
For demonstration purposes, the accumulation of irreversible strain in the titanium alloy VT6 (Russian analog of Ti-6Al-4V) is analysed. Three types of phenomenological models of plastic ratcheting are considered. They are the Armstrong-Frederick model as well as the first and the second Ohno-Wang models. Based on real data, a new rule of isotropic hardening is proposed for greater accuracy of simulation.  The ability of the sensitivity analysis to determine reliable and unreliable parameters is demonstrated. The plausibility of the new method is checked by alternative approaches, like the consideration of correlation matrices and validation of identified parameters on ``unseen'' data. A relation between pathological error sensitivity and overparametrization is established.
\end{abstract}

\section{Introduction}

Accurate simulation of the stress-strain curves in a broad range of loading scenarios is essential for fatigue strengths assessments. Proper models of elasto-plasticity are back-bone of many advanced modelling approaches, including purely phenomenological models of continuum damage mechanics and microstructure-motivated models. Moreover, some simplified engineering methods allow for assessments of fatigue life, based on the parameters of the stress-strain hysteresis loops \cite{Collins1993, Yang2005, Zhu2017}.

Both phenomenological and microstructure-based models are calibrated against experimental data. In general, the low sensitivity of material parameters to experimental errors means that the problem of parameter identification is stable, and the identified parameters can be used to solve practical problems. However, the high sensitivity, also called pathological sensitivity, means that even a slight noise in experimental data may cause essential changes in the identified parameters. This sensitivity renders the obtained parameters unreliable and useless.
In this regard, the following questions arise: (i) Are the material constants pathologically dependent on the measurement errors? (ii) How to quantify the sensitivity of a certain identification procedure? (iii) How to estimate the sensitivity?
To answer these questions, we further develop the methodology from \cite{ShutovKaygorodtsevaZAMM2019, ShutovKaygorodtsevaActaMech2020}, used to study the sensitivity
of material parameters respective to the noise in experimental data.
A step-by-step algorithm is presented allowing to estimate the sensitivity in a mechanically reasonable way. As a part of the approach, a new mechanics-based metric in the space of material parameters is developed especially for ratcheting-related applications. The sensitivity of parameters is then evaluated concerning this metric.

As a demonstration example, we analyse the ratcheting of the titanium alloy VT6. The main focus is on accurate modelling of the thermo-mechanical response at mid-life and the corresponding sensitivities to measurement errors.
In the case of uniaxial cyclic loading, considered here, nonlinear kinematic hardening must be accounted for due to the presence of the Bauschinger effect.\footnote{Dealing with a general non-proportional loading, advanced models of directional distortional hardening are needed \cite{Francois2001}, \cite{ShutovIhlemann2012}, \cite{ShutovPanhans2011}. Moreover,
simplified J2 yield conditions can be replaced by their anisotropic counterparts \cite{Adamus2015}.} Three models are considered in this study: the Armstrong-Frederick model (AF), the first, and the second Ohno-Wang models (OW-I and OW-II). The isotropic hardening is modelled according to a new rule, based on the accumulated total strain and the Odqvist parameter.

Since we focus on the initial stage and mid-life, ductile damage is not incorporated into the considered models. However, this can be made based on principles from \cite{Lemaintre1984, Bruenig2008, ShutovSilbermannInhelamm2015, Bartel2016}.
Application of ductile damage models to ratcheting of metals is found in \cite{Kang2009, Surmiri2019}.

To determine the optimal set of material parameters, we implement a nested identification procedure. It is a combination of the gradient-free Nelder-Mead method and the gradient-based Levenberg-Marquardt method. Owing to the new rule of isotropic hardening, all models enable accurate simulations in agreement with the experimental data. The simulation accuracy increases with increasing number of parameters. Validation of the parameters and models is carried out by comparing the simulated and ``unseen'' experimental data regarding strain accumulation and dissipative heating.
Moreover, for various identification problems correlations between unknown parameters are computed. Typically, high correlations appear in overparametrized models.

The plausibility of the newly proposed error-sensitivity analysis is checked by comparison with other methods. Validation of material parameters on ``unseen'' data and correlations between parameters agree with the results of the sensitivity analysis.

The paper is organized as follows. Section 2 presents the general algorithm of the sensitivity analysis; the main ingredients and ideas are shown. Four sufficient criteria of overparametrized model are introduced. In Section 3, new models of ratcheting accounting for cyclic mechanical loading are presented. The novelty lies in the advanced rule of isotropic hardening.  The heat conduction equation is developed using the first law of thermodynamics.
The experimental results for the alloy VT6, corresponding parameter identification protocols, and validation of parameters are show in Section 4.
Section 5 is devoted to the sensitivity analysis for the specific models of ratcheting. The implemented stochastic model of noise and the mechanics-based metric are presented, and the parameter sensitivities are estimated, depending on the complexity of the model. The occurrence of overparametrized models is exposed. Section 6 discusses the results, and final conclusions are presented in Section 7.

\section{Error sensitivity and overparametrization}

\subsection{Basic steps of sensitivity analysis}

To formulate the general algorithm, we consider an arbitrary material model of ratcheting. Assume that the model contains $n$ real-valued material parameters, subject to identification. By $\vec{p} \in \mathbb{R}^n$, we denote the unknown parameter vector. The available experimental data are packed into the vector $\overrightarrow{Exp} \in \mathbb{R}^{N_{exp}}$, $N_{exp} \geq n$. For ratcheting-related applications, such measured data can be total strains, strain amplitudes, displacements at individual points or even discretized displacement fields. Let $\overrightarrow{Mod}(\vec{p})$ be the corresponding model prediction of $\overrightarrow{Exp}$.

\textbf{Identification.}
Following the standard approach \cite{Beck2007}, the error functional $\Phi$ is built and the required parameter vector $\vec{p}^{\ \ast}$ is its minimizer:
\begin{equation}\label{TargetFunctionGeneral}
\Phi(\vec{p}) = ( \overrightarrow{Exp} -\overrightarrow{Mod}(\vec{p})) \cdot \mathbf{W}  \cdot (\overrightarrow{Exp} -\overrightarrow{Mod}(\vec{p})), \quad \vec{p}^{\ \ast} = \text{argmin} \Phi(\vec{p}) .
\end{equation}
Here, $\mathbf{W}$ is a fixed, symmetric, positive-definite weighting matrix. In the simplest case when all the experimental data are uncorrelated and equally important, one typically takes the identity matrix in place of $\mathbf{W}$ \cite{Beck2007, ShutovKaygorodtsevaZAMM2019}.

\textbf{Distance between parameter sets.} For sensitivity studies a reasonable metric is needed in the space of material parameters. In the following sections, a mechanics-based distance is introduced such that $\text{dist}(\vec{p}_1, \vec{p}_2)$ is a deviation of parameter sets  $\vec{p}_1$ and $\vec{p}_2$ from each other.

\textbf{Stochastic model of noise.}
To account for eventual presence of experimental errors, we introduce a stochastic model of experimental noise. Assuming that the noise is additive \cite{Beck2007, Harth2004, Harth2007}, the real experimental data $\overrightarrow{Exp}$ are replaced by the noisy data $\overrightarrow{NoisyData} = \overrightarrow{Exp} + \overrightarrow{Noise}$. Here, $\overrightarrow{Noise}$ is a random vector in $\mathbb{R}^{N_{exp}}$; its distribution is defined by the stochastic model, discussed later.

\textbf{Monte Carlo computations.}
Let $N_{\text{noise}}$ be a sufficiently large number of draws of noisy data and the $j$th draw be denoted as $\overrightarrow{Noise}^{(j)}$. For each draw we consider a new optimization problem with the error function
\begin{equation}\label{TargetFunctionGeneralNoisy}
\Phi^{\text{noisy}}(\vec{p}) = ( \overrightarrow{Exp} + \overrightarrow{Noise}^{(j)} -\overrightarrow{Mod}(\vec{p})) \cdot \mathbf{W}  \cdot (\overrightarrow{Exp} + \overrightarrow{Noise}^{(j)}  -\overrightarrow{Mod}(\vec{p})), \quad \vec{p}^{\ (j)} = \text{argmin} \Phi^{\text{noisy}}(\vec{p}) .
\end{equation}
A highly efficient way of solving this optimization problem is presented in Appendix A.
The set of vectors $\vec{p}^{\ (j)} $, $j = 1,2,..., N_{\text{noise}}$ is called the ``cloud of parameters''. The center of the cloud is the average of all $\vec{p}^{\ (j)}$ and the size of the cloud is the average distance to the center:
\begin{equation}\label{SizeParamCloudGeneral}
\vec{p}^{\ \text{center}} = \frac{1}{N_{\text{noise}}} \sum_{j=1}^{N_{\text{noise}}} \vec{p}^{\ (j)}, \quad
CloudSize = \frac{1}{N_{\text{noise}}} \sum_{j=1}^{N_{\text{noise}}} \text{dist}(\vec{p}^{\ \text{center}}, \vec{p}^{\ (j)}).
\end{equation}

If we assume that 
the probability density function (PDF) of noise is symmetric with respect to zero ($\text{PDF}(\vec{x}) = \text{PDF}(-\vec{x})$ for all $\vec{x} \in \mathbb{R}^{N_{exp}}$), then $\vec{p}^{\ \ast}$ is located roughly at the center of the parameter cloud. In this case the definition of the $CloudSize$ simplifies to
\begin{equation}\label{SizeCloudSimplified}
CloudSize = \frac{1}{N_{\text{noise}}} \sum_{j=1}^{N_{\text{noise}}} \text{dist}(\vec{p}^{\ \ast}, \vec{p}^{\ (j)}).
\end{equation}
The $CloudSize$ is the measure of the sensitivity of identified material parameters. Small $CloudSize$ for realistic noise indicates that the sensitivity is low and the strategy is stable. If $CloudSize$ is finite even for vanishing noise, then the identification protocol is unreliable and the dependence on measurement errors is pathological. Such a pathological dependence is characteristic of overparametrized models.

For reproducibility of results and faster convergence of Monte Carlo computations, the quasi-Monte Carlo method is implemented in the current study (cf. \cite{ Niederreiter1978, ShutovKaygorodtsevaActaMech2020}).
The quasi-Monte Carlo method differs from the classical in using a low-discrepancy sequence of random numbers.
In the current paper, the Sobol sequence is implemented \cite{Sobol1967}.

\subsection{Criteria of overparametrization}

\begin{figure}\centering
\scalebox{0.8}{\includegraphics{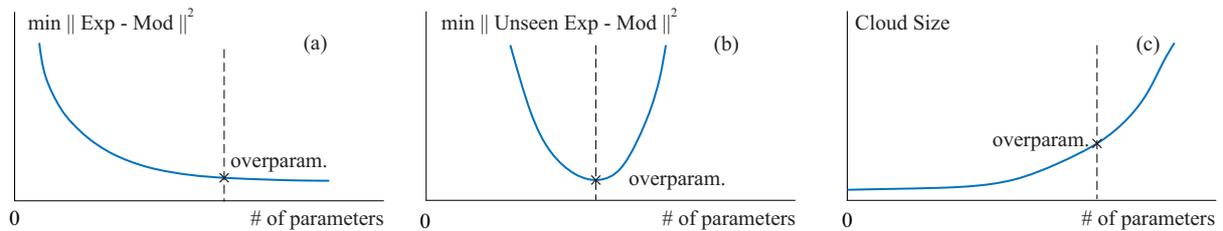}}
\caption{Signs of overparamterized material models for increasing number of  parameters:
inadequate gain in accuracy (left), deteriorating predictive capabilities tested on ``unseen'' data (middle), and unacceptably high sensitivity of parameters to measurement errors (right).}
\label{figSketchOverparam}
\end{figure}

We say that the model is overparametrized if one of the following criteria is satisfied:
\begin{itemize}
    \item[I]  Virtually no gain in accuracy occurs when increasing the number of material parameters (Fig. \ref{figSketchOverparam}(left)).
    \item[II] The validation of the model on ``unseen'' data shows deteriorating  predictive capabilities as the number of parameters increases (Fig. \ref{figSketchOverparam}(middle)).
    \item[III] There is a significant correlation among parameters.
\end{itemize}

In the current paper, we suggest the fourth criterion:
\begin{itemize}
    \item[IV]  The sensitivity of material parameters to experimental error becomes too high  (Fig. \ref{figSketchOverparam}(right)).
\end{itemize}

In the following sections, the usefulness of this new criterion is analysed. Toward that end, the basic steps of the sensitivity analysis are demonstrated on concrete examples. The applicability of the algorithm is demonstrated and the plausibility of the results is assessed by comparison with criteria I, II, and III.

\section{Material models}

For simplicity, the small strain framework is implemented. However, the developed models can be generalized to large strains using the methodology from
\cite{Lion2000, ShutovKreissig2008, Vladimirov2008, ShutovLarichkin2017}. For phenomenological material description, three types of combined isotropic-kinematic hardening models are used here: the models of Armstrong-Frederick (AF), the first, and the second Ohno-Wang models (OW-I and OW-II). To control the complexity of each modelling approach, we introduce two, three, and four rheological branches. These branches are the following: the rate-independent Maxwell body for the AF-model, elsto-plastic model of Prandtl-Reuss for the OW-I-model, and a modified Maxwell body for the OW-II-model.

All the models share the same kinematics. First, the total strain tensor $\boldsymbol{\varepsilon}$ is additively decomposed into the thermal part  $\boldsymbol{\varepsilon}_{\theta}$ and the mechanical part $\boldsymbol{\varepsilon}_{\text{m}}$:
\begin{equation}
\boldsymbol{\varepsilon} = \boldsymbol{\varepsilon}_{\theta} +  \boldsymbol{\varepsilon}_{\text{m}}.
\end{equation}
The temperature-induced strain is purely volumetric; it is given by $\boldsymbol{\varepsilon}_{\theta} = \frac{1}{3}\alpha({\theta} - {\theta}_0)\cdot \mathbf{1}$, where $\alpha$ is the (volumetric) thermal expansion coefficient and ${\theta}_0$ is the reference temperature.
The mechanical strain is decomposed into the elastic part $\boldsymbol{\varepsilon}_{\text{e}}$ and the inelastic part $\boldsymbol{\varepsilon}_{\text{i}}$, such that
$\boldsymbol{\varepsilon}_{\text{m}} =  \boldsymbol{\varepsilon}_{\text{e}} + \boldsymbol{\varepsilon}_{\text{i}}$.
Let $N_{\text{branches}} = 2, 3, 4$  be the number of rheological branches. For each branch we introduce conservative ($\boldsymbol{\varepsilon}_{l \text{e}}$) and dissipative  ($\boldsymbol{\varepsilon}_{l \text{i}}$) components of strain, such that
\begin{equation}\label{COnserDissipy}
\boldsymbol{\varepsilon}_{\text{i}} = \boldsymbol{\varepsilon}_{l \text{e}} + \boldsymbol{\varepsilon}_{l \text{i}}, \quad \text{for all} \quad l = 1,...,N_{\text{branches}}.
\end{equation}
The Helmholz free energy per unit mass is computed as:
\begin{equation}\label{freeEnergy}
\Psi =  \Psi_{\text{e}}(\boldsymbol{\varepsilon}_{\text{e}}) + \sum\limits_{l=1}^{N_{\text{branches}}} \Psi_{\text{kin} l }(\boldsymbol{\varepsilon}_{l \text{e}}) + \Psi_{\theta}(\theta),
\end{equation}
\begin{equation}\label{Psi_e}
\rho \Psi_{\text{e}}(\boldsymbol{\varepsilon}_{\text{e}})  = \frac{k}{2} (\text{tr} \boldsymbol{\varepsilon}_{\text{e}})^2 + \mu \boldsymbol{\varepsilon}_{\text{e}}^{\text{D}}:\boldsymbol{\varepsilon}_{\text{e}}^{\text{D}},
\end{equation}
\begin{equation}\label{Psi_kin}
\rho \Psi_{\text{kin} l }(\boldsymbol{\varepsilon}_{l \text{e}}) = \frac{c_l}{2}\boldsymbol{\varepsilon}_{l \text{e}}^{\text{D}}:\boldsymbol{\varepsilon}_{l \text{e}}^{\text{D}}, \quad l = 1,...,N_{\text{branches}},
\end{equation}
\begin{equation}\label{Psi_theta}
 \rho \Psi_{\theta}(\theta) = -c_{\theta 0} \Big(\theta \ln{\frac{\theta}{\theta_0}} - (\theta-\theta_0)\Big).
\end{equation}
Here, $k>0$ and $\mu >0$ are the bulk and shear moduli; $c_l >0$ is the stiffness of the substructure, described by branch $l$; $c_{\theta 0}$ is material's heat capacity per unit mass. Note that alternative assumptions regarding energy storage $\Psi_{\theta}$ are also possible, cf. \cite{Bartel2016}.
The stress tensor $\boldsymbol{\sigma}$ is computed through Hooke's law; the backstresses $\mathbf{X}_l$ in branches $l = 1,...,N_{\text{branches}}$ are linear isotropic functions of  $\boldsymbol{\varepsilon}_{l \text{e}} = \boldsymbol{\varepsilon}_{\text{i}}-\boldsymbol{\varepsilon}_{l\text{i}}$:
\begin{equation}\label{Hooke}
\boldsymbol{\sigma} = \frac{\partial \Psi_{\text{e}}}{\partial \boldsymbol{\varepsilon}_{\text{e}}} = k \text{tr} (\boldsymbol{\varepsilon}_m-\boldsymbol{\varepsilon}_{\text{i}})\mathbf{1} + 2\mu(\boldsymbol{\varepsilon}_m-\boldsymbol{\varepsilon}_{\text{i}})^{\text{D}},
\end{equation}
\begin{equation}\label{BackStress}
\mathbf{X}_l = \frac{\partial \Psi_{\text{kin} l }}{\partial \boldsymbol{\varepsilon}_{l\text{e}}} = c_l (\boldsymbol{\varepsilon}_{\text{i}}-\boldsymbol{\varepsilon}_{l\text{i}})^{\text{D}}, \quad l=1,\ldots,N_{\text{branches}},
\end{equation}
where $\mathbf{A}^{\text{D}} = \mathbf{A} - \frac{1}{3} \text{tr} (\mathbf{A}) \mathbf{1}$.
The effective stress $\boldsymbol{\sigma}^{\text{eff}}$ equals
\begin{equation}
\boldsymbol{\sigma}^{\text{eff}} = \boldsymbol{\sigma}  - \sum\limits_{l=1}^{N_{\text{branches}}}\mathbf{X}_l.
\end{equation}
In a similar fashion, we compute the entropy per unit mass as:
\begin{equation}\label{Enthropy}
\zeta = -\frac{\partial \Psi }{\partial \theta}|_{\boldsymbol{\varepsilon} = const}  = -\frac{\partial \Psi_{\theta} }{\partial \theta} + \frac{\alpha}{3\rho} \text{tr} \boldsymbol{\sigma} = \frac{c_{\theta 0}}{\rho} \ln{\frac{\theta}{\theta_0}}+ \frac{\alpha}{3\rho} \text{tr} \boldsymbol{\sigma}.
\end{equation}
In the following, relations \eqref{Hooke}, \eqref{BackStress}, and \eqref{Enthropy}, along with implemented evolution equations, will be sufficient for thermodynamic consistency of the models.

Let $K >0$ be the initial uni-axial yield stress. By $R \in \mathbb{R}$ we denote the isotropic hardening. Then the viscous overstress $f$ and the inelastic strain rate  $\lambda_{\text{i}} = || \dot{ \boldsymbol{\varepsilon}_{\text{i}} } ||$ are computed using
\begin{equation}\label{Perzyna}
f :=  ||(\boldsymbol{\sigma}^{\text{eff}} )^{\text{D}}|| - \sqrt{\frac{2}{3}}(K+R), \quad
\lambda_{\text{i}} = \frac{1}{\eta}\Big\langle\frac{f}{f_0}\Big\rangle^{m_{\text{Perzyna}}}, \quad \langle x \rangle := \max(x,0).
\end{equation}
Here, $\eta >0$ and $m_{\text{Perzyna}} >0$ are parameters of Perzyna's viscosity law; $f_0=1$ MPa.
In the previous paper \cite{Kaygorodtseva2020} it was shown that the classical Voce's law of isotropic hardening is inappropriate for the ratcheting of the VT6 alloy. Therefore, the evolution of the isotropic hardening $R$ is modelled by a  \textit{new law}, which accounts both for hardening and softening:
\begin{equation}\label{HardeningLaw}
R =  \gamma s - \beta s_{\varepsilon}, \quad  \text{where} \quad \dot{s} = \sqrt{\frac{2}{3}} || \dot{\boldsymbol{\varepsilon}}_{\text{i}} ||  , \quad \dot{s}_{\varepsilon} =\sqrt{\frac{2}{3}} || \dot{\boldsymbol{\varepsilon}}^{\text{D}} ||.
\end{equation}
In this law, $s$ is the accumulated plastic arc-length (also known as Odqvist parameter), $s_{\varepsilon}$ is the accumulated total strain, $\gamma \in \mathbb{R}$ and $\beta$ are material constants.
For simplicity, we assume that the initial state of the material is isotropic. Therefore, the following initial conditions are used:
\begin{equation}\label{InitialConditions}
\boldsymbol{\varepsilon}_{\text{i}}|_{t=0} = \boldsymbol{\varepsilon}_{1\text{i}}|_{t=0} = \ldots = \boldsymbol{\varepsilon}_{4\text{i}}|_{t=0} = \mathbf{0}, \quad s|_{t=0} = s_{\text{d}}|_{t=0} = 0.
\end{equation}
In all the models, the global flow rule governs the inelastic strain rate $\boldsymbol{\varepsilon}_{\text{i}}$ according to the normality rule:
\begin{equation}\label{NormalityRule}
\dot{ \boldsymbol{\varepsilon}_{\text{i}} } =
\lambda_{\text{i}} \frac{(\boldsymbol{\sigma}^{\text{eff}} )^{\text{D}}}{||(\boldsymbol{\sigma}^{\text{eff}} )^{\text{D}}||}.
\end{equation}

\textbf{Armstrong-Frederick hardening.} For  the model of AF-type the constitutive equations of each branch correspond to a rate-independent Maxwell model:
\begin{equation}\label{NormalityRuleAF}
\dot{\boldsymbol{\varepsilon}}_{l\text{i}} = \lambda_{\text{i}} \ \varkappa_l \ \mathbf{X}_l, \quad l = 1,...,N_{\text{branches}}.
\end{equation}
Here, $\varkappa_l \geq 0$ is the material parameter controlling the saturation of the backstress $\mathbf{X}_l$.

\textbf{Remark.} Differentiating \eqref{BackStress} with respect to time and using \eqref{NormalityRuleAF}, we obtain a well-known form of the Armstrong-Frederick equation:
\begin{equation}\label{StressBasedFormula}
\dot{\textbf{X}}_l = c_l (\dot{\boldsymbol{\varepsilon}}_{\text{i}} - \dot{\boldsymbol{\varepsilon}}_{l\text{i}})^{\text{D}} = c_l  \big( \dot{\boldsymbol{\varepsilon}}_{\text{i}} -  \varkappa_l \lambda_{\text{i}} \mathbf{X}_l \big).
\end{equation}
Although \eqref{NormalityRuleAF} and \eqref{StressBasedFormula} are equivalent, we prefer using strain-based relation \eqref{NormalityRuleAF} rather than the stress-based formula \eqref{StressBasedFormula}.

\textbf{First Ohno-Wang hardening.}
For the OW-I model, each branch corresponds to elastic-perfectly plastic body, also known as the Prandtl-Reuss body. The constitutive equations are
\begin{equation}\label{NormalityRuleOW}
\dot{\boldsymbol{\varepsilon}}_{l\text{i}} = \lambda_{l\text{i}} \frac{\mathbf{X}_l}{||\mathbf{X}_l||}, \quad \lambda_{l\text{i}} \geq 0 , \quad ||\mathbf{X}_l||\leq\sqrt{\frac{2}{3}} r_l.
\end{equation}
Here $\lambda_{l\text{i}}$ is the inelastic strain rate in the $l$th branch, $r_l> 0$ is the corresponding yield stress.
Moreover, the Kuhn-Tucker conditions must be satisfied:
\begin{equation}\label{Kuhn-Tucker}
\lambda_{l\text{i}} \ \Big(||\mathbf{X}_l|| - \sqrt{\frac{2}{3}} r_l \Big) = 0.
\end{equation}
The historical stress-based evolution law (cf. \cite{OhnoWang1993}) can be obtained by combining \eqref{BackStress}, \eqref{NormalityRuleOW}, and \eqref{Kuhn-Tucker}.

\textbf{Second Ohno-Wang hardening.}
For the OW-II model we use
\begin{equation}\label{NormalityRuleOW-II}
\dot{\boldsymbol{\varepsilon}}_{l\text{i}} =
\Big(\sqrt{\frac{2}{3}} \frac{||\mathbf{X}_l||}{r_l}\Big)^m \Big\langle\dot{\boldsymbol{\varepsilon}}_{\text{i}} : \frac{\mathbf{X}_l}{||\mathbf{X}_l||}\Big\rangle \frac{\mathbf{X}_l}{||\mathbf{X}_l||}.
\end{equation}
Here, the parameter $r_l$ has a similar meaning as in the OW-I model.
The exponent $m$ governs the degree of nonlinearity.

\textbf{Remark.} Differentiating \eqref{BackStress} with respect to time and substituting \eqref{NormalityRuleOW-II} into the result, we obtain the historical stress-based form of the evolution equation (cf. \cite{OhnoWang1993}):
\begin{equation}
\dot{\textbf{X}}_l =  c_l \Bigg(\dot{\boldsymbol{\varepsilon}}_{\text{i}} -  \lambda_{\text{i}}
\Big(\sqrt{\frac{2}{3}} \frac{||\mathbf{X}_l||}{r_l}\Big)^m \Big\langle  \frac{(\boldsymbol{\sigma}^{\text{eff}} )^{\text{D}}}{||(\boldsymbol{\sigma}^{\text{eff}} )^{\text{D}}||}: \frac{\mathbf{X}_l}{||\mathbf{X}_l||}\Big\rangle \frac{\mathbf{X}_l}{||\mathbf{X}_l||} \Bigg).
\end{equation}
Again, we prefer dealing with the strain-based equation \eqref{NormalityRuleOW-II}.

\textbf{Remark.} Another popular ratcheting model was proposed by Abdel-Karim and Ohno in \cite{AbdelKarimOhno2000}. This model combines dynamic recovery terms from the AF and OW-I models. However, it is not considered in the current study, since it contains too many material parameters.

Let $\delta_\text{i}$ be the energy dissipation per unit mass.  According to the second law of thermodynamics (cf. \cite{Haupt2013}), the Clausius-Duhem inequality must be satisfied for all possible thermo-mechanical processes:
\begin{equation}\label{ClausiusDuhem}
\delta_\text{i} = \frac{1}{\rho} \boldsymbol{\sigma}: \dot{\boldsymbol{\varepsilon}} - \dot{\Psi} - \dot{\theta}\xi \geq 0.
\end{equation}
Using \eqref{Psi_e}, \eqref{Psi_kin},  \eqref{Hooke}, and \eqref{Enthropy} we obtain the reduced form of the mechanical dissipation:
\begin{equation}\label{DissipatReduced}
\delta_\text{i} = \frac{1}{\rho} \boldsymbol{\sigma}^{\text{eff}} : \dot{\boldsymbol{\varepsilon}}_{\text{i}} + \frac{1}{\rho} \sum_{l=1}^{N_{\text{branches}}} \mathbf{X}_l:\dot{\boldsymbol{\varepsilon}}_{l\text{i}} \geq 0.
\end{equation}
Substituting evolution equations into the reduced form, we can check that \eqref{DissipatReduced} is indeed satisfied:
\begin{equation}\label{CheckDissipat}
\boldsymbol{\sigma}^{\text{eff}} : \dot{\boldsymbol{\varepsilon}}_{\text{i}} \geq 0, \quad
\mathbf{X}_l:\dot{\boldsymbol{\varepsilon}}_{l\text{i}} \geq 0.
\end{equation}
The non-negativity of the dissipation indicates that \emph{all the models are thermodynamically consistent}.
In addition, we postulate the first law of thermodynamics in the local form:
\begin{equation}\label{1LawOfThermodynamics}
\theta \dot{\zeta} = \delta_\text{i} - \frac{1}{\rho}\text{div} \boldsymbol{q} + r.
\end{equation}
Here, $\boldsymbol{q}$ is the heat flux vector, $r$ represents local heat sources per unit mass.
To simulate the heating of the sample in the gauge area, we implement the following simplified heat-exchange scheme (cf. \cite{ShutovIhlemann2011})
\begin{equation}\label{SimplHeatExch}
 - \frac{1}{\rho}\text{div} q + r = \omega (\theta - \theta_0).
\end{equation}
Here, $\theta$ is the sample temperature in the gage area; the temperature of the surrounding medium is assumed to be equal to the reference temperature $\theta_0$; $\omega$ is the heat-exchange coefficient, depending on the shape and dimensions of the sample as well as on the heat conduction properties of involved components.
Now we differentiate \eqref{Enthropy} with respect to time to obtain the rate of entropy:
\begin{equation}
 \dot{\zeta} = \frac{c_{\theta 0}}{\rho} \frac{\dot{\theta}}{\theta} + \frac{\alpha k}{\rho} \text{tr} \dot{\boldsymbol{\varepsilon}}  - \frac{\alpha^2}{\rho} k  \dot{\theta}.
\end{equation}
Multiplying both sides with $\theta$, we have
\begin{equation}
\theta \dot{\zeta} = \frac{c_{\theta 0}}{\rho} \dot{\theta} + \frac{\alpha k}{\rho} \text{tr} \dot{\boldsymbol{\varepsilon}} \theta - \frac{\alpha^2}{\rho} k  \theta \dot{\theta}.
\end{equation}
Using the heat capacity parameter $ c_{\theta} = \frac{c_{\theta 0}}{\rho} - \frac{\alpha^2}{\rho} k  \theta$ we arrive at
\begin{equation}
\theta \dot{\zeta} = c_{\theta} \dot{\theta} + \frac{\alpha k}{\rho} \text{tr} \dot{\boldsymbol{\varepsilon}} \theta.
\end{equation}
Substituting this into \eqref{1LawOfThermodynamics}, we obtain the heat conduction equation:
\begin{equation}
c_{\theta} \dot{\theta} = -\frac{\alpha k \theta}{\rho} \text{tr} \dot{\boldsymbol{\varepsilon}} + \delta_\text{i} - \omega ({\theta} - {\theta}_0).
\end{equation}
On the right side of this equation, the first term is responsible for the thermoelastic effect, the second term describes dissipation-induced heating, and the third term is the heat flux from the sample into the environment.

The thermal part is introduced into the modelling framework to validate the model by available experimental results on sample heating. In this work, the temperature dependence of the constants is neglected, since in the considered application the temperature increase is insignificant.

\section{Parameter identification for VT6}

\subsection{Experimental data}

\begin{figure}\centering
\scalebox{0.8}{\includegraphics{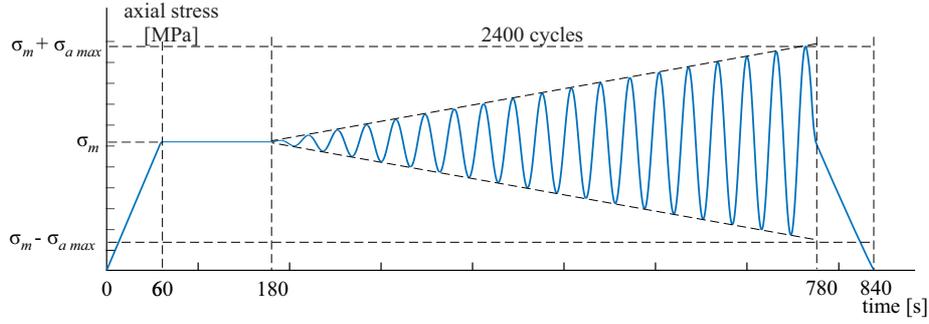}}
\caption{Loading program implemented in the experiment \cite{Kaygorodtseva2020}.}
\label{figLoadingProgram}
\end{figure}

We use experimental data from  \cite{Kaygorodtseva2020} on the ratcheting of samples from the titanium alloy VT6. In each test, the loading program consists of four stages (Fig. \ref{figLoadingProgram}):
quasi-static monotonic loading, holding under constant stress, harmonic cyclic loading with linearly increasing stress amplitude, and unloading. The third stage of each test contains 2400 stress-controlled cycles.
During this stage, the mean stress $\sigma_m$ is held fixed and the stress amplitude is monotonically increasing; the maximum stress amplitude in each test is denoted as $\sigma_{a \ max}$.

\begin{figure}\centering
\scalebox{0.8}{\includegraphics{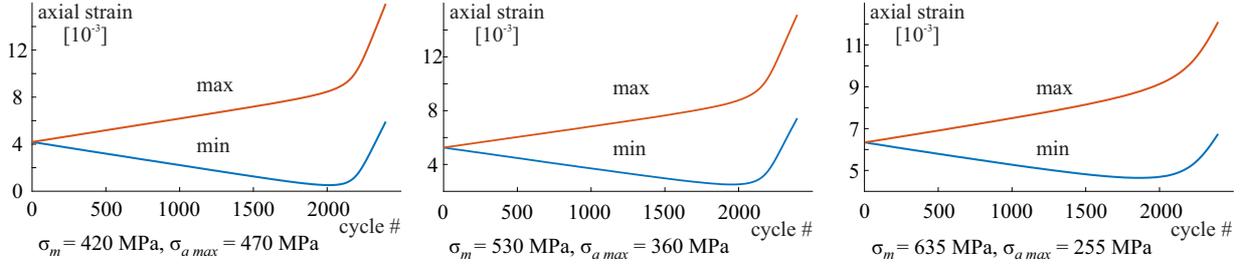}}
\caption{Experimental data from \cite{Kaygorodtseva2020} on the ratcheting of the VT6 alloy.}
\label{figExpData}
\end{figure}

Here we consider three tests where the maximum and the minimum axial strains at each cycle are recorded; for 2400 cycles this makes 4800 values per test.
 Tests with $\sigma_m = 420$ MPa, $\sigma_{a \ max} = 470$ MPa (Fig. \ref{figExpData}(left)) and $\sigma_m = 635$ MPa, $\sigma_{a \ max} = 255$ MPa (Fig. \ref{figExpData}(right)) are used for parameter identification. The test with $\sigma_m = 530$ MPa, $\sigma_{a \ max} = 360$ MPa (Fig. \ref{figExpData}(middle)) is reserved for validation. The maximum achievable stress for all tests is the same: it equals $\sigma_m + \sigma_{a \ max} = 890$ MPa.

\subsection{Identification procedure}

The viscous effects are neglected here, so we put $\eta \rightarrow 0$, $m_{\text{Perzyna}}=1$. The elastic constants of VT6 are fixed to $k = 98,037$ MPa, $\mu = 37,593$ MPa. AF-type models contain 7, 9, and 11 unknown parameters for variants with two, three and four Maxwell branches, respectively.
The number of free parameters in models of OW-I-type is always less by one than in corresponding AF-models.
Namely, to ensure that the material retains its carrying capacity,
one of the branches needs to remain purely elastic all the time. Therefore,
the corresponding local yield stress $r_l$ has to be sufficiently large. Since this parameter is undefinable using the available data, it is fixed to a pre-defined large value and it is excluded from the identification procedure. In contrast, each of the OW-II models contains one additional parameter, namely, the exponent $m$.

In this section, we describe the identification procedure for models with $N_{\text{branches}} = 4$. For convenience, we introduce the following notation:
\begin{equation}
\vec{p}_c = (\gamma, \beta, c_1, c_2, c_3, c_4), \quad
    \vec{p}_K  =
\begin{cases}
    (\varkappa_1, \varkappa_2, \varkappa_3, \varkappa_4, K) \quad  \  \text{for AF},  \\
    (r_1, r_2, r_3, K) \quad \quad \ \quad \ \text{for OW-I}, \\
    (r_1, r_2, r_3, r_4, K, m) \quad \text{for OW-II} .
\end{cases}
\end{equation}
To identify parameters, tests with $\sigma_m = 420$ MPa and $\sigma_m = 635$ MPa are used. Denote by $\overrightarrow{Exp} \in \mathbb{R}^{9600}$ the vector of the experimental data. For each model, the vector of the model response  $\overrightarrow{Mod}(\vec{p}_c, \vec{p}_K)$ contains the simulation results, corresponding to $\overrightarrow{Exp}$.
Following the previously discussed standard procedure we build an error function $\Phi(\vec{p}_c, \vec{p}_K)$ as the average deviation of the simulation results from the experimental data:
\begin{equation}\label{TargetFunction}
\Phi(\vec{p}_c, \vec{p}_K) = ( \overrightarrow{Exp} -\overrightarrow{Mod}(\vec{p}_c, \vec{p}_K)) \cdot (\overrightarrow{Exp} -\overrightarrow{Mod}(\vec{p}_c, \vec{p}_K)).
\end{equation}

\textbf{Remark.} In \eqref{TargetFunction}, all the experimental data enter the error function  \eqref{TargetFunction} with the same weight. However, in some applications it is advisable to provide weighting factors or even weighting matrices \cite{Beck2007, ShutovKaygorodtsevaZAMM2019}.

The standard identification procedure minimizes the error function $\Phi(\vec{p}_c, \vec{p}_K)$. However, due to the large number of material parameters, the error function $\Phi$ may exhibit numerous local minima. Therefore, optimization algorithms may provide a bad fit of simulation to experiment. To solve this issue, we use a nested identification procedure. It consists of internal and external optimization problems.
The internal optimization problem is as follows. For a fixed set $\vec{p}_K$, the parameters $\vec{p}_c$ are identified from the following partial minimization:
\begin{equation}\label{InternalOptimization}
\vec{P}_c (\vec{p}_K) = \mathop{\text{argmin}}_{\vec{p}_c} \ \Phi(\vec{p}_c, \vec{p}_K).
\end{equation}
The external optimization makes use of the internal optimization:
\begin{equation}\label{ExternalOptimization}
\vec{p}_K = \mathop{\text{argmin}}_{\vec{p}_K} \ \Phi(\vec{P}_c(\vec{p}_K), \vec{p}_K), \quad \vec{p}_c = \vec{P}_c(\vec{p}_K).
\end{equation}

The parameter identification procedure is implemented using the gradient-free Nelder-Mead method \cite{NelderMead1965}. After the nested identification procedure is complete, the obtained set of material parameters is refined using the gradient-based Levenberg-Marquardt method \cite{LevenbergMarquardt2005}. Within the Levenberg-Marquardt step, general identification is carried out where all parameters are identified simultaneously. Such a refinement is necessary to make the gradient of the error function equal to zero. As will be seen from the following, zero gradient is essential for efficient algorithms implemented in the sensitivity analysis.

\subsection{Results of identification and validation}

\begin{figure}\centering
\scalebox{0.8}{\includegraphics{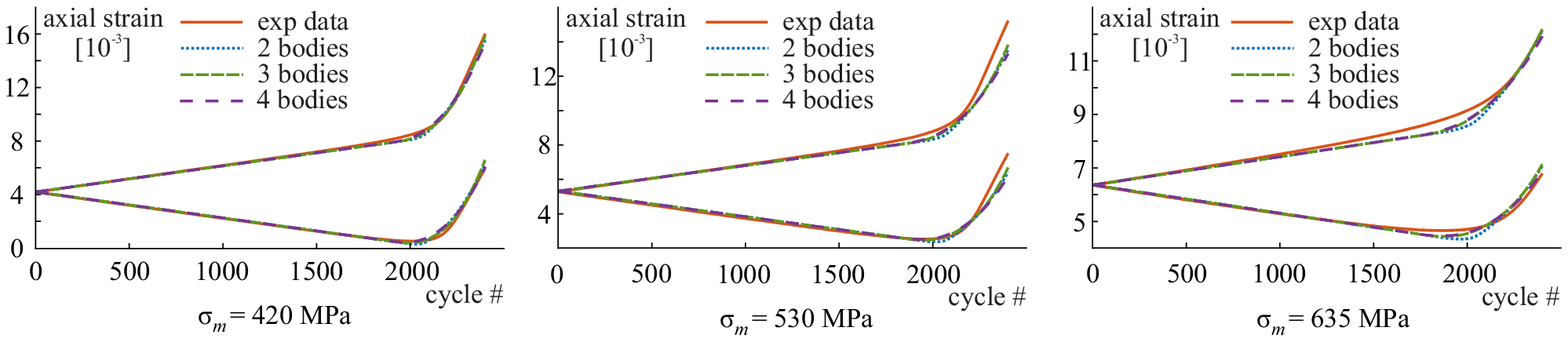}}
\caption{Experimental data and simulation results by AF models for various number of branches (rheological bodies).}
\label{figExpVsModAF}
\end{figure}

\begin{figure}\centering
\scalebox{0.8}{\includegraphics{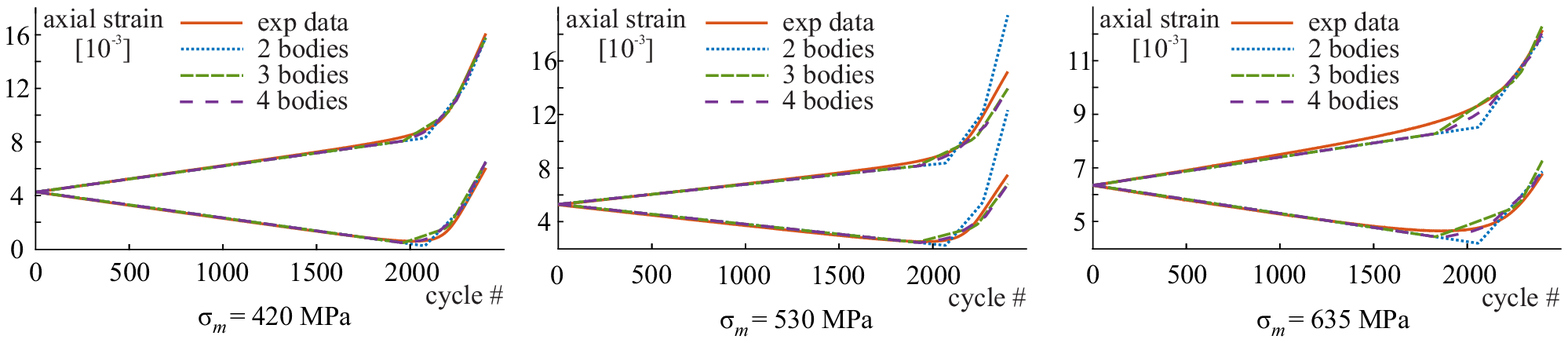}}
\caption{Experimental data and simulation results by OW-I models for various number of branches (rheological bodies).}
\label{figExpVsModOW-I}
\end{figure}

\begin{figure}\centering
\scalebox{0.8}{\includegraphics{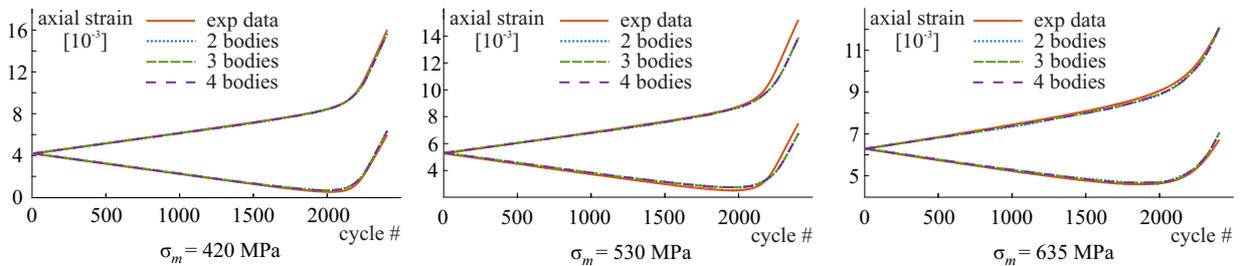}}
\caption{Experimental data and simulation results by OW-II models for various number of branches (rheological bodies).}
\label{figExpVsModOW-II}
\end{figure}

The identification results for AF, OW-I, and OW-II are given in Tables \ref{ParamSetAF}, \ref{ParamSetOW-I} and \ref{ParamSetOW-II},  respectively.
The best fit of AF models to experiment is shown in Fig. \ref{figExpVsModAF}(left and right); best identification results for OW-I models are shown in Fig. \ref{figExpVsModOW-I}(left and right); best results for OW-II are in Fig. \ref{figExpVsModOW-II}(left and right).
For validation, the cyclic test with $\sigma_m = 530$ MPa is used, see Fig. \ref{figExpVsModAF}(middle), Fig. \ref{figExpVsModOW-I}(middle) and Fig. \ref{figExpVsModOW-II}(middle).

\begin{table}[h]
\centering
\caption{Material parameters for AF models}
\begin{tabular}{c c c c c c c c}  \hline
    \multicolumn{7}{c}{conservative parameters}\\ \hline
    $N_{\text{branches}}$&$\gamma$[MPa]& $c_1$[MPa]& $c_2$[MPa]& $c_3$[MPa]& $c_4$[MPa] \\\hline
    2  &8094.2 & 12005 & 143832 & - & - \\
    3  &5736.0 & 7777.4 & 18789 & 109793 & - \\
    4  &4176.2 & 4294.6 & 8232.5 & 21724 & 117736  \\ \hline
    \multicolumn{7}{c}{dissipative parameters}\\ \hline
    $N_{\text{branches}}$ & $\beta$[-] & $\varkappa_1$ [1/MPa] & $\varkappa_2$ [1/MPa] & $\varkappa_3$ [1/MPa]& $\varkappa_4$ [1/MPa] & $K$[MPa] \\ \hline
    2  &3.7978 & 0.0360 & 0.0906 & - & - & 862.86  \\
    3  &3.5277 & 0.0352 & 0.0527 & 0.0866 & - & 847.26  \\
    4  &4.0121 & 0.0227 & 0.0366 & 0.0646 & 0.0797 & 846.73  \\   \hline
\end{tabular} \\
\label{ParamSetAF}
\end{table}

\begin{table}[h]
\centering
\caption{Material parameters for OW-I models}
\begin{tabular}{c c c c c c c c}  \hline
    \multicolumn{7}{c}{conservative parameters}\\\hline
    $N_{\text{branches}}$&$\gamma$[MPa]& $c_1$[MPa]& $c_2$[MPa]& $c_3$[MPa]& $c_4$[MPa] \\ \hline
    2  &4527.7 & 7329.5 & 4714.3 & - & - \\
    3  &2109.3 & 10004 & 17306 & 7962.7 & - \\
    4  &6475.8 & 14915 & 19164 & 9673.5 & 3765.9  \\ \hline
    \multicolumn{7}{c}{dissipative parameters}\\ \hline
    $N_{\text{branches}}$&$\beta$[-]& $r_1$[MPa]& $r_2$ [MPa] & $r_3$[MPa]& $r_4$[MPa]& $K$[MPa] \\ \hline
    2  &4.0919 & 30.702& $\infty $ & - & -  & 884.69  \\
    3  &3.7940 & 22.206 & 31.518& $\infty $ & - &  852.41  \\
    4  &3.8239 & 7.6188 & 17.099 & 29.030 & $\infty $&  856.30  \\   \hline
\end{tabular} \\
\label{ParamSetOW-I}
\end{table}

\begin{table}[h]
\centering
\caption{Material parameters for OW-II models }
\begin{tabular}{c c c c c c c c c}  \hline
    \multicolumn{8}{c}{conservative parameters}\\\hline
    $N_{\text{branches}}$&$\gamma$[MPa]& $c_1$[MPa]& $c_2$[MPa]& $c_3$[MPa]& $c_4$[MPa] \\ \hline
    2  &8957.3 & 214914 & 18441 & - & - \\
    3  &8785.4 & 498547 & 10857 & 70184 & - \\
    4  & 8805.0 & 140688 & 11377 & 443284 & 74932  \\ \hline
    \multicolumn{7}{c}{dissipative parameters}\\ \hline
    $N_{\text{branches}}$&$\beta$[-]& $r_1$[MPa]& $r_2$ [MPa] & $r_3$[MPa]& $r_4$[MPa]& $K$[MPa] & $m$[-] \\ \hline
    2  &3.6190 & 101.26 & 39.032 & - & -  & 757.30 & 2.9817  \\
    3  &3.6194 & 99.053 & 27.750 & 58.154 & - &  713.53 & 3.0173  \\
    4  & 3.6195 & 18.853 & 28.665 & 86.949 & 59.896 &  704.02 & 3.0490 \\   \hline
\end{tabular} \\
\label{ParamSetOW-II}
\end{table}

All the considered models show a good correspondence between the simulation and the experiment. Naturally, as the number of rheological branches grows large, the accuracy increases. However, the validation of AF models on ``unseen'' data shows slightly decreasing accuracy for larger $N_{\text{branches}}$. Poor validation on ``unseen'' data indicates that AF models may be overparametrized for $N_{\text{branches}}=4$.

\begin{figure}\centering
\scalebox{0.8}{\includegraphics{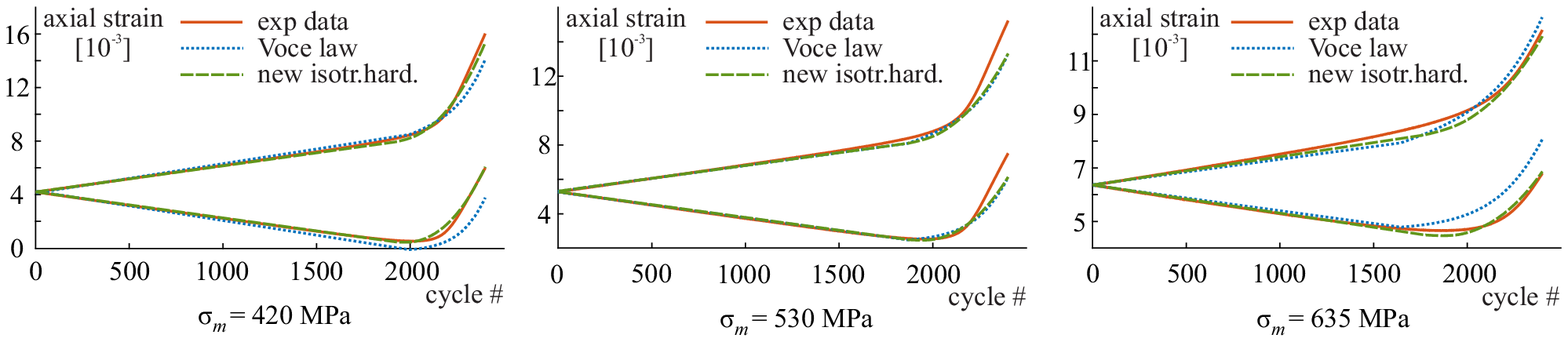}}
\caption{Best possible fit using the classical Voce rule \eqref{ClassicalVoce} and the new rule of isotropic hardening \eqref{HardeningLaw}.}
\label{figVoceVsNewRule}
\end{figure}

In many theoretical and applied studies, the classical Voce rule of isotropic hardening is used
\begin{equation}\label{ClassicalVoce}
R(s) = \frac{p_1}{p_2}(1 - \text{e}^{\displaystyle -p_2 s}),
\end{equation} where $p_1$ and $p_2$ are material parameters and $s$ is the Odqvist parameter. In \cite{Kaygorodtseva2020} the applicability of the classical rule \eqref{ClassicalVoce} was studied.  The Odqvist parameter $s$ was discovered to be inappropriate for the accurate description of the isotropic hardening in VT6. This problem is a motivation for the new rule of isotropic hardening presented in equation \eqref{HardeningLaw}. The presented simulation results show that the new rule is much more accurate than the classical Voce rule (Fig. \ref{figVoceVsNewRule}), although it contains the same number of material parameters.

\begin{figure}\centering
\scalebox{0.8}{\includegraphics{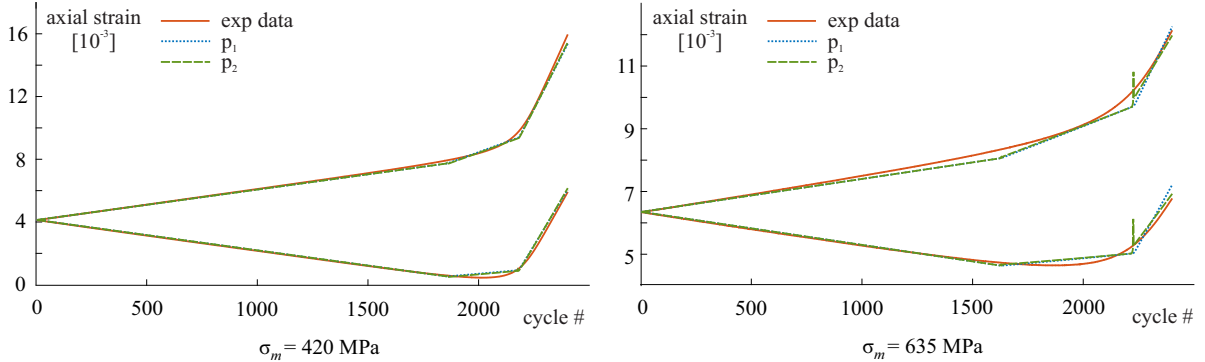}}
\caption{Experimental data and simulation results by OW-I model obtained for $\vec{p}_1$ and $\vec{p}_2$; $N_{\text{branches}} = 2$.}
\label{figModOWDifferP}
\end{figure}

Special attention is required for OW-I models. In Table \ref{ParamSetOWDifferP} the vector of parameters $\vec{p}_1$ corresponds to the identification procedure, described earlier, and $\vec{p}_2$ is obtained by a single iteration of the Gauss-Newton method starting from $\vec{p}_1$. Although the gradient of the error function $\Phi$ is virtually zero at $\vec{p}_1$, a single step of the Gauss-Newton method yields a huge increment of the parameter vector (Table \ref{ParamSetOWDifferP}).
In Fig. \ref{figModOWDifferP} we show the experimental data and simulation results for the OW-I model with $N_{\text{branches}} = 2$, corresponding to parameter vectors $\vec{p}_1$ and $\vec{p}_2$; the simulation results for $\vec{p}_2$ are obtained by the linearization of the model response near  $\vec{p}_1$. The deviation of the simulation results from the experiment is nearly the same for both simulations. This example shows that the identification procedure for the OW-I model with this set of experimental data is unstable; it cannot provide reliable parameters for use in practical applications.  The ill-posedness of the identification problem for the models of OW-I type is also visible from correlation matrices.
The procedure for computing the correlation matrix and its results are summarized in Appendix B.
As seen from Tables \ref{CorrOW-I2}, \ref{CorrOW-I3}, and \ref{CorrOW-I4}, the correlation coefficient between some of the parameters are precisely equal to one.  Loosely speaking, the reason for high correlation and unreliable optimization is as follows: the optimization algorithm ``does not know'' which values of the micro yield stresses $r_l$ should be taken for the best correspondence between simulation and experiment.
Since the identification procedure for the OW-I model is clearly unreliable, its error-sensitivity is not studied any further.

\begin{table}[h]
\centering
\caption{Material parameters of the OW-I model.}
\begin{tabular}{c c c c c c c c }  \hline
        &$\gamma$[MPa]& $c_1$[MPa]& $c_2$[MPa]& $\beta$[-]& $r_1$[MPa] &$r_2$[MPa] & $K$[MPa]\\ \hline
    $p_1$ &0.0926 & 67,285 & 10,040 & 3.681 & 84.088 & $\infty$ & 825.43 \\
    $p_2$ &18,001,000 & 70,563 & 14,692,000 & 4.508 & 91.289 & $\infty$ & 832.01 \\
  \hline
\end{tabular} \\
\label{ParamSetOWDifferP}
\end{table}

\begin{figure}\centering
\scalebox{0.8}{\includegraphics{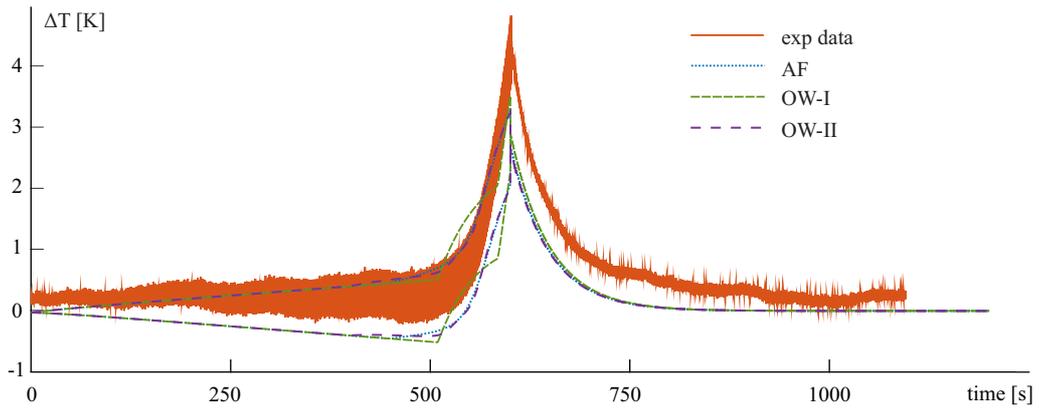}}
\caption{Experimental data on temperature evolution and simulation results, $N_{\text{branches}} = 2$.}
\label{figTemperature2}
\end{figure}

\begin{figure}\centering
\scalebox{0.8}{\includegraphics{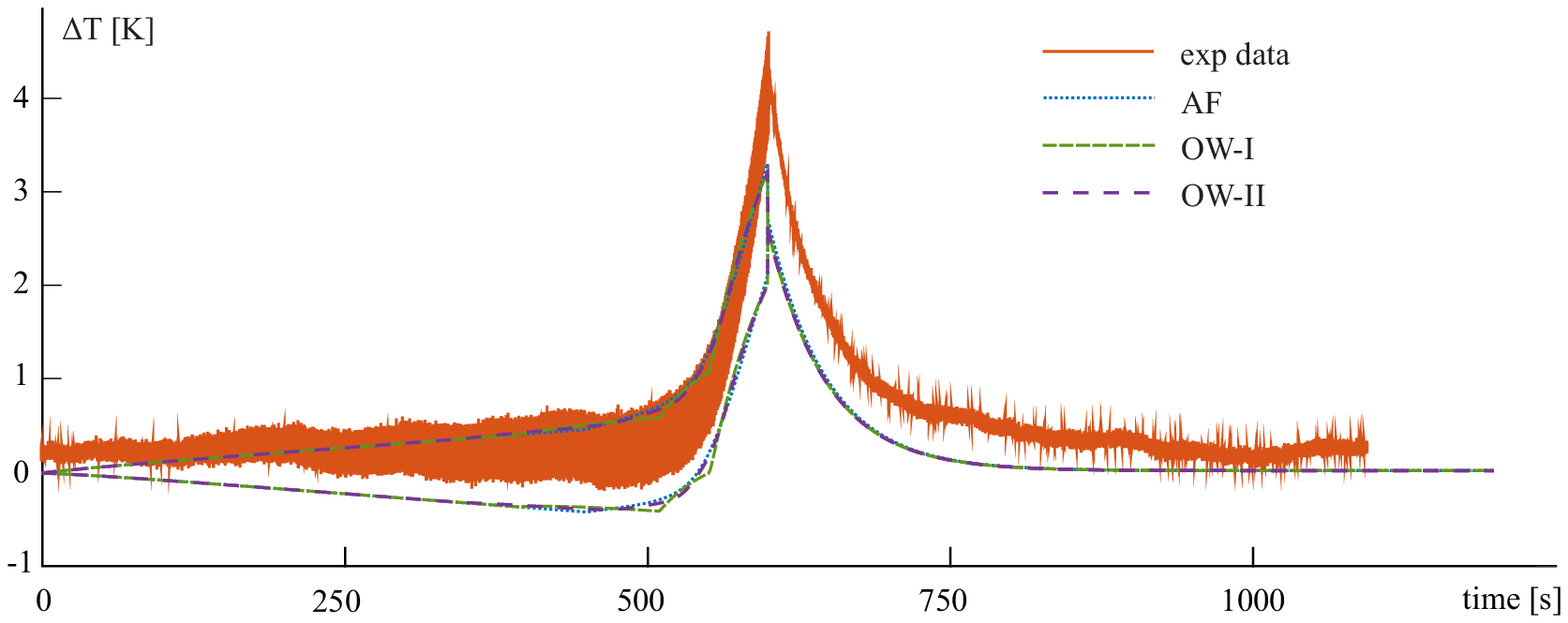}}
\caption{Experimental data on temperature evolution and simulation results, $N_{\text{branches}} = 3$.}
\label{figTemperature3}
\end{figure}

\begin{figure}\centering
\scalebox{0.8}{\includegraphics{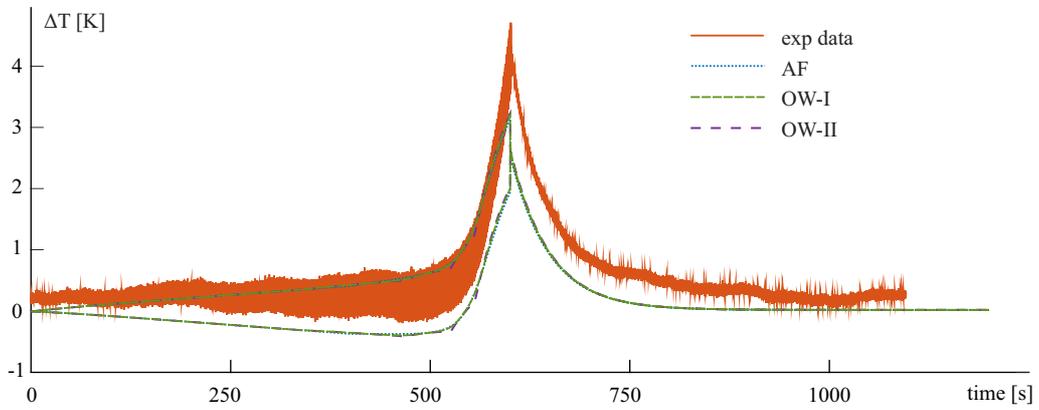}}
\caption{Experimental data on temperature evolution and simulation results, $N_{\text{branches}} = 4$.}
\label{figTemperature4}
\end{figure}

For the final validation of the developed models, we use experimental data on the temperature evolution of the sample in the gauge area.
Note that the experimental data on the temperature increase were not used during the identification. The following temperature-related parameters are used: the volumetric thermal expansion $\alpha = 1.59 \cdot 10^{-5}$ [1/K], the heat capacity per unit volume $c_{\theta 0} / \rho = 1.2058 \ [\text{J}/(\text{kg} \cdot \text{K})]$, the mass density $\rho = 4,550 \ [\text{kg} / \text{m}^3]$. The heat exchange coefficient $w =  2.5\cdot 10^{-2} \ [\text{J} / (\text{s} \cdot \text{kg} \cdot \text{K}) ]$ is chosen to provide a realistic fit during cooling of the sample.
As is seen from Figs. \ref{figTemperature2},  \ref{figTemperature3}, and \ref{figTemperature4}, AF-type models are the most realistic regarding the temperature evolution. All the simulation results agree with the experimental data. The simulation reproduces the thermoelastic effect and the dissipation-induced heating. OW-I and OW-II models differ essentially from the experimental data for
$N_{\text{branches}}=2$, but the results are plausible for $N_{\text{branches}}=4$.

\section{Sensitivity analysis}

\subsection{Stochastic model of noise}

As mentioned in Section 2, the actual experimental data contain errors. Following \cite{Harth2004, Harth2007, Beck2007}, the errors are additive. Therefore, the vector of noisy data $\overrightarrow{NoisyData}$ is the sum of given experimental data $\overrightarrow{Exp}$ and the noise $\overrightarrow{Noise}$: $\overrightarrow{NoisyData} = \overrightarrow{Exp} + \overrightarrow{Noise}$. In this work a simple stochastic model is implemented:
\begin{equation}\label{StochModel}
NoisyData_i = Exp_i + \sum_{k=1}^{20} \sigma_k \cdot \text{Mode}_k (t_i), \quad \text{Mode}_k (t) = \sin\Big(k \pi\frac{t}{T}\Big), \quad \sigma_k \in \mathcal{N}(0,\sigma^2).
\end{equation}
Here, $\sigma_k \in \mathcal{N}(0,\sigma^2)$ are independent, normally distributed random values with zero mean and the variance $\sigma^2$; $\text{Mode}_k (t)$ is the $k$th mode of noise with $t \in [0, T]$.
Since the measured strains are non-dimensional, so is $\sigma$. In the current computations, the standard deviation is $\sigma = 10^{-6}$.

The probability density function of noise is symmetric: $\text{PDF}(\vec{x}) = \text{PDF}(-\vec{x})$ for all $\vec{x} \in \mathbb{R}^{N_{exp}}$. Thus, simplified definition of the $CloudSize$ is valid and equation \eqref{SizeCloudSimplified} is used instead of  $\eqref{SizeParamCloudGeneral}_2$.

\begin{figure}\centering
\scalebox{1}{\includegraphics{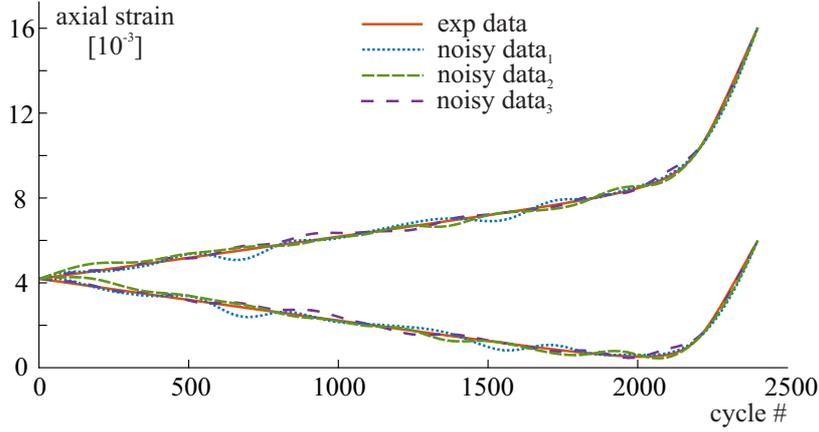}}
\caption{Illustration of experimental data and three draws of the noisy data according to the stochastic model \eqref{StochModel}.}
\label{figNoisyData}
\end{figure}

The reader interested in other stochastic models of noise, is referred to \cite{ShutovKaygorodtsevaZAMM2019, AvrilEtAl2004, Grediac2004, Seibert2000}. Studies dealing with correlated noise are   \cite{Harth2004, Harth2007, Beck2007}.
In many practical situations the real stochastic model of noise is unknown. In that case, the maximum entropy principle is helpful in creating realistic models \cite{Soize2017, Wolszczak2019}.

As mentioned in Section 2, quasi Monte Carlo computations are based on Sobol's sequence instead of pseudo-random numbers. A step-by-step procedure for computing coefficients $\sigma_k$ which appear in \eqref{StochModel} is described in Appendix C.

\subsection{Mechanics-based metric for ratcheting models}

\begin{figure}\centering
\scalebox{0.8}{\includegraphics{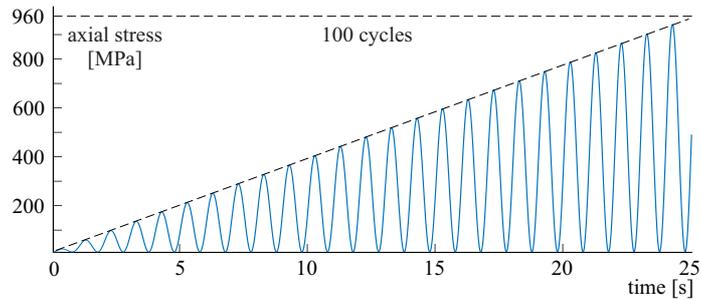}}
\caption{Stress-controlled loading program used to compute the mechanics-based metric \eqref{PhysBasedDist}.}
\label{figLoadingProgramForMetric}
\end{figure}

To estimate the sensitivity of material parameters to the experimental errors, we introduce a new mechanics-based metric. This metric gives the distance between two sets of material parameters. The metric idea is to simulate the same stress-controlled ratcheting process using two sets of parameters. The distance is the maximum discrepancy between the strain trajectories:
\begin{equation}\label{PhysBasedDist}
\text{dist}(\vec{p}^{\ (1)}, \vec{p}^{\ (2)}) := \max\limits_{t \in [0,T_{\text{metric}}]}
| \varepsilon_{11}(t,\vec{p}^{\ (1)}) - \varepsilon_{11}(t,\vec{p}^{\ (2)})  |.
\end{equation}
Here, $T_{\text{metric}}$ is the overall duration of the stress-controlled ratcheting process; $\varepsilon_{11}$ is the solution of the uniaxial ratcheting problem. To be definite, the corresponding stress-controlled loading corresponds to cyclic loading with $\sigma_{\text{min}} = 0$ and monotonically increasing $\sigma_{\text{max}}$ (Fig. \ref{figLoadingProgramForMetric}).
The stress amplitude should be chosen such that the accumulated plastic strain $s$ would not exceed the values of $s$ obtained in real experiments.

\textbf{Remark.} The metric defined by \eqref{PhysBasedDist} is similar to the one introduced for plasticity models in \cite{ShutovKaygorodtsevaZAMM2019}. The main difference is as follows: For the plasticity models, the metric is based on the assumption that the material model obtains the strain history as input and provides the stress history as output. Thus, the plasticity-related metric is computed in terms of stresses \cite{ShutovKaygorodtsevaZAMM2019}. In ratcheting-related applications, however, the model obtains the stress history as input and provides the total strain as output. Therefore, the proposed distance \eqref{PhysBasedDist} is a non-dimensional number.

The advantages of the new mechanics-based metric are the following:
\begin{itemize}
    \item the metric is invariant under reparamterization of the material model;
    \item the metric accounts for changes in those parameters which have a substantial impact on the mechanical response; less important parameters are automatically disregarded;
    \item the metric is independent of constants, used to obtain non-dimensional parameters (cf. \cite{ShutovKaygorodtsevaZAMM2019});
    \item the application-related metric is obtained by choosing specific stress history.
\end{itemize}

\subsection{Size of the parameter cloud}

Let $\vec{p}^{\ \ast}$ be the set of optimal parameters, obtained from the minimization of the error function $\Phi$. For given experimental data, the set $\vec{p}^{\ \ast}$ is deterministic. Recall that the sensitivity analysis is based on the (quasi) Monte Carlo method.
Let $N_{\text{noise}}$ be the number of draws of noisy data. Each draw is made according to the stochastic model \eqref{StochModel}.
Let $\vec{p}^{\ (j)}$ be the set of parameters for the $j$th draw of noisy experimental data, $j = 1, 2, ..., N_{\text{noise}}$.
Each $\vec{p}^{\ (j)}$ is obtained by the computationally efficient procedure in Appendix A.

\textbf{Remark.} Reasonable $\vec{p}^{\ (j)}$ are computed only when the noise-free optimal solution $\vec{p}^{\ \ast}$ is highly accurate. At $\vec{p}^{\ \ast}$, the gradient of the original error function $\Phi$ must be as close to zero as possible. This is the reason why the refined optimization procedure was implemented in Section 4 to find $\vec{p}^{\ \ast}$.

Recall that the size of the parameter cloud is
\begin{equation}\label{SizeParamCloud}
CloudSize = \frac{1}{N_{\text{noise}}} \sum_{j=1}^{N_{\text{noise}}} \text{dist}(\vec{p}^{\ \ast}, \vec{p}^{\ (j)}).
\end{equation}
Here,  $N_{\text{noise}} = 10,000$. It is important that the mechanics-based metric is used here. Fast computation of $\text{dist}(\vec{p}^{\ \ast}, \vec{p}^{\ (j)})$ is explained in Appendix D.

\begin{table}
\centering
\caption{Sizes of the parameter clouds for AF and OW-II models in terms of the mechanics-based metric. }
\begin{tabular}{ p{2.0cm} p{2.5cm}p{2.5cm} p{2.5cm} }
\hline
   model:   & $N_{\text{branches}} = 2 $  & $N_{\text{branches}} = 3 $  & $N_{\text{branches}} = 4 $  \\ \hline
   AF    & 0.000284  & 0.00114   & 0.00342   \\
   OW-II & 0.0000142
   & 0.000299 & 0.000440   \\
  \hline

\end{tabular}
\label{TableCloudSize}
\end{table}

Table \ref{TableCloudSize} shows the error-sensitivity of AF and OW-II models based on the available experimental data. Clearly, the OW-II models are calibrated much more reliably than the AF models.

\textbf{Comparison of overparametrization criteria.}
The validation on ``unseen'' data shows that predictive capabilities of AF-type models are deteriorating as the number of rheological branches $N_{\text{branches}}$ increases. According to criterion II from Section 2.2, this indicates that the AF-model with $N_{\text{branches}}=4$ is overparamterized.
Next, criterion III of an overparametrized model is the appearance of high correlation among the parameters. Indeed, dealing with the AF-models with large  $N_{\text{branches}}$, the correlation is too high (Tables \ref{CorrAF2}, \ref{CorrAF3},  \ref{CorrAF4}). Dealing with the AF-model with $N_{\text{branches}} = 4$, the correlation between $\gamma$ and $c_1$ equals 1.0, and the correlation between $K$ and $\varkappa_3$ ranges up to 0.9999, warning of overparametrization.

With increasing number of rheological branches $N_{\text{branches}}$, $CloudSize$ is growing monotonically. Thus, calibration of models with a larger number of parameters is more sensitive to measurement errors than for their simpler counterparts. For certain level of noise, the $CloudSize$ may become unacceptably large (cf. criterion IV). Thus, for the AF models, the commonly used criteria are consistent with the new criterion IV; there is a clear relation between overparamterization and pathological error-sensitivity.

A  different situation is observed for OW-II models. Even for $N_{\text{branches}} = 4 $ the correlation among the parameters is smaller than 1.0 (Tables \ref{CorrOW-II2}, \ref{CorrOW-II3}, and \ref{CorrOW-II4}). Besides, for increasing $N_{\text{branches}}$ the predictive capabilities tested on ``unseen'' data are not deteriorating (Fig. \ref{figExpVsModOW-II}). Moreover, the cloud sizes are much smaller than for the AF models (Tables \ref{TableCloudSize}). Therefore, according to the basic criteria, the OW-II model is not overparametrized even for $N_{\text{branches}} = 4 $. This is a striking difference with the OW-I model, whose calibration is clearly unreliable. Again, for OW-II models, the criterion IV is also consistent with the classical criteria of overparametrization.

\section{Discussion}

In phenomenological material modelling, the standard procedure is as follows. First, the model and numerical algorithm are created, followed by calibration and validation of the model against experimental data. In this study we propose an additional step, namely, the analysis of the sensitivity of identified material parameters with respect to errors contained in experimental data.
Based on quasi Monte Carlo computations, the procedure gives insights into the error propagation through the simulation cycle.
In choosing among different protocols of parameter identification, the preference should be given to the robust.
Reliable identification strategies show low sensitivity of identified material parameters; this automatically rules out overparametrized models.

Following the classical protocol of parameter identification, the problem is reduced to minimization of the error functional $\Phi$. As error-resistant solution of optimization problems is imperative in many engineering applications, various alternative approaches were also developed \cite{Wolszczak2019}. The advantage of the method advocated in the current study lies in its simplicity and practical use.

We show that for increasing number of parameters, the modelling becomes more accurate, but the  error-sensitivity increases (Table \ref{TableCloudSize}). Thus, a conflict between accuracy and stability appears. A similar conflict was previously reported in \cite{ShutovKaygorodtsevaZAMM2019} for plasticity-related applications.

The general drawback of conventional plasticity models,  is the kink of the stress-strain curve at the elastic-plastic boundary. One elegant way to solve this problem is introduction of subloading yield surfaces \cite{Hashiguchi1989, Hashiguchi2017}. A simpler way to smoothen the stress-strain curve, used by many, is to
take a very large stiffness $c_k$ in one of the rheological branches ($c_k \gg \mu$). The undesired side effect is that an unacceptably large correlation appears among other parameters.
In the AF models, large $c_k$ triggers a correlation between $K$ and
$\varkappa_k$, see Tables \ref{CorrAF2}, \ref{CorrAF3}, and \ref{CorrAF4}.
For OW-I and OW-II models, $K$ and $r_k$ correlate, see Tabsles \ref{CorrOW-I2} --- \ref{CorrOW-II4}.  Thus, dealing with $c_k \gg \mu$, special regularization is needed to avoid the undesired correlation.

Interestingly, when $N_{\text{branches}}=4$ all three models perform equally well in describing the temperature evolution of the sample (Fig. \ref{figTemperature4}). This similarity is due to convergence of the stress-strain curves to the same limit for large number of rheological branches. Since the dissipation-induced heating depends on the area of hysteresis loops, the temperature evolution in the tested models is nearly identical.

In the tests considered, the models of OW-II type are more stable to experimental errors than the AF-models. This stability is especially unexpected since OW-II models contain more parameters than the AF-models. Detailed analysis and explanation of this effect is a subject of a separate study.

\section{Conclusion}

The paper presents a procedure for the sensitivity analysis of ratcheting models. The procedure is exemplified by actual experimental data and by various types of kinematic hardening rules, like AF, OW-I, and OW-II.
The plausibility of the sensitivity studies is assessed by comparison with alternative approaches, like the use of correlation matrix and validation on ``unseen'' data. The main conclusions are as follows:
\begin{itemize}
    \item For models of ratcheting, a new mechanics-based metric is presented, allowing to measure the distance between sets of material parameters.
    \item A computationally efficient quasi Monte Carlo procedure is used to estimate the stability of the identified parameters; a large number of draws can be taken ($N_{\text{noise}} \geq 10,000$).
    \item A new rule of isotropic hardening is proposed. Dealing with the VT6 alloy, it is more accurate than the classical Voce rule based on the accumulated plastic arc-length.
    \item  For increasing number of parameters, the size of the parameter cloud grows large. This indicates that there is a relation between pathological error-sensitivity and overparametrization.
    \item The results of sensitivity analysis are consistent with computations of the correlation matrices: Unreliable identification procedures with large error-sensitivities are also characterized by strong correlations between some of parameters. The results are also consistent with other analysis methods, like validation on ``unseen''  data.
    \item In the considered example, the calibration of OW-II models is more reliable than the calibration of AF and OW-I models.
\end{itemize}

The presented ideas behind the sensitivity analysis are rather general.
The core of the method is based on Monte Carlo computations, a general tool, suitable even for integration of irregular functions in multiple dimensions \cite{Benedetti2020}.
It is promising for analysis of parameter identification involving a large amount of noisy data, like data provided by digital image correlation in tests with a heterogeneous stress state.

\textbf{Acknowledgments.} The authors are thankful for stimulating discussions of the Monte Carlo method with Dr. I.N. Medvedev (Novosibirsk, Russia). We are also thankful to V.I. Kapustin and K.V. Zakharchenko (Novosibirsk, Russia) for providing experimental data on temperature evolution. The research was supported by the Russian Science Foundation (project number 19-19-00126).

\textbf{Compliance with ethical standards}

\textbf{Conflict of interest} The authors declare that they have no
conflict of interest.

\section*{Appendix A: Fast computation of $\vec{p}^{\ (j)}$}

We discuss a quick computation of the parameter vectors $\vec{p}^{\ (j)} \in \mathbb{R}^{n}$, corresponding to $j$th draws of noisy data. The procedure is the same as in \cite{ShutovKaygorodtsevaActaMech2020}. Recall that $\overrightarrow{Exp} \in \mathbb{R}^{N_{\text{exp}}}$ is the vector of available experimental data, $\overrightarrow{Mod}(\vec{p}) \in \mathbb{R}^{N_{\text{exp}}}$ is the corresponding modelling response, $\vec{p} = (\vec{p}_c, \vec{p}_K) \in \mathbb{R}^{n}$ is the vector of unknown material parameters. Within the sensitivity analysis, the actual experimental data are replaced by the noisy data $\overrightarrow{Exp} + \overrightarrow{Noise}$.
The optimal set of parameters corresponding to noise-free data is denoted as $\vec{p}^{\ \ast}$. The Jacobian of the model response at $\vec{p}^{\ \ast}$ is the operator
\begin{equation}\label{JacobDefinit}
\mathbf{J} := \frac{\partial \overrightarrow{Mod}(\vec{p})} {\partial \vec{p}}|_{\vec{p}^{\ \ast}} \in \mathbb{R}^{N_{\text{exp}} \times n}.
\end{equation}
Assuming only small changes in parameters, we linearize the model response near $\vec{p}^{\ \ast}$:
\begin{equation}
\overrightarrow{Mod}^{lin}(\vec{p}) := \overrightarrow{Mod}(\vec{p}^{\ \ast}) + \mathbf{J}(\vec{p} - \vec{p}^{\ \ast}).
\end{equation}
Now, the parameter set $\vec{p}^{\ (j)}$, $j = 1, 2, ..., N_{\text{noise}}$ is the minimizer of the  error function for noisy data
\begin{equation}
\Phi^{\text{noisy}}(\vec{p}) := \overrightarrow{Resid}^{\text{T}} \ \overrightarrow{Resid},
\end{equation}
\begin{equation}
\overrightarrow{Resid} := \overrightarrow{Exp} + \overrightarrow{Noise} - \overrightarrow{Mod}^{lin} =
\overrightarrow{Exp} + \overrightarrow{Noise} -  \overrightarrow{Mod}(\vec{p}^{\ \ast})  - \mathbf{J}(\vec{p} - \vec{p}^{\ \ast}).
\end{equation}
Abbreviate by $\overrightarrow{A}$ the following vector:
\begin{equation}\label{NotationA}
\overrightarrow{A} :=  \overrightarrow{\text{Exp}} + \overrightarrow{\text{Noise}} - \overrightarrow{\text{Mod}(\vec{p}^{\ \ast})} - \mathbf{J} \ \vec{p}^{\ \ast} .
\end{equation}
Then the error function is a quadratic form of $\vec{p}$, given by
\begin{equation}
\Phi^{\text{noisy}}(\vec{p}) = \big(\overrightarrow{A} - \mathbf{J} \vec{p}\big)^{\text{T}} \big(\overrightarrow{A} -  \mathbf{J} \vec{p}\big).
\end{equation}
Its derivative with respect to $\vec{p}$ is a linear function of the unknown parameter vector $\vec{p}$
\begin{equation}
\frac{\displaystyle \partial \Phi^{\text{noisy}}(\vec{p})}{\displaystyle \partial \vec{p} } = - 2
 \big( \overrightarrow{A}  -  \mathbf{J} \vec{p} \big)^{\text{T}}   \ \mathbf{J}.
\end{equation}
The stationarity condition $\frac{\displaystyle \partial \Phi^{\text{noisy}}(\vec{p})}{\displaystyle \partial \vec{p} } = 0$ yields a system of linear algebraic equations with respect to $\vec{p}$. Then the analytical solution is
\begin{equation}\label{AnalytSolution}
\vec{p}^{\ (j)} = \big( \mathbf{J}^{\text{T}} \mathbf{J}  \big)^{-1} \big( \mathbf{J} \big)^{\text{T}} \ \overrightarrow{A}.
\end{equation}
In fact, this semi-analytical solution represents a single iteration of the Gauss-Newton method \cite{WrightNocedal1999}.

Unfortunately, due to matrix multiplication, the condition number of  $\mathbf{J}^{\text{T}} \mathbf{J}$ can be very large. This effect may falsify the results of \eqref{AnalytSolution}. To resolve this problem, $\mathbf{Q} \mathbf{R}$ decomposition of $\mathbf{J}$ should be implemented:
\begin{equation}\label{QRDecomp}
\mathbf{J} = \mathbf{Q} \ \mathbf{R} \in \mathbb{R}^{N_{\text{exp}} \times n}, \quad
\mathbf{Q} \in \mathbb{R}^{N_{\text{exp}} \times n}, \quad
\mathbf{Q}^{\text{T}} \mathbf{Q} = id_{\mathbb{R}^{n}} \in \mathbb{R}^{n \times n}, \quad
\mathbf{R} \in \mathbb{R}^{n \times n}.
\end{equation}
Here, $\mathbf{R}$ is upper triangular. Substituting this into \eqref{AnalytSolution}, a  more robust formula is obtained:
\begin{equation}\label{RobsutSolution}
\vec{p}^{\ (j)} = \big( \mathbf{R}^{\text{T}} \mathbf{R}  \big)^{-1} \mathbf{R}^{\text{T}} \mathbf{Q}^{\text{T}}  \ \overrightarrow{A}.
\end{equation}
Since the matrix $\big( \mathbf{R}^{\text{T}} \mathbf{R}  \big)^{-1} \mathbf{R}^{\text{T}} \mathbf{Q}^{\text{T}}$ is pre-computed, the parameter cloud is evaluated extremely efficiently even for a large number of draws ($N_{\text{draws}} \geq 10,000$).

\section*{Appendix B: Correlation matrices}

Let $\mathbf{J} $ be the Jacobian, defined in \eqref{JacobDefinit}. The correlation matrix $\mathbf{Corr} \in \mathbb{R}^{n \times n}$ is defined as follows \cite{Benedix2000}:
\begin{equation}
\mathbf{Corr}_{i j} = \mathbf{P}_{i j}/ \sqrt{ \mathbf{P}_{i i} \mathbf{P}_{j j} }, \quad \text{where} \quad \mathbf{P} = \mathbf{J}^{\text{T}} \ \mathbf{J} \in \mathbb{R}^{n \times n}.
\end{equation}
We say that there is a strong correlation between parameters $p_i$ and $p_j$, if $\mathbf{Corr}_{i j} \approx  \pm 1$.
In that case, a slight change in $p_i$  can be counteracted by a change in $p_j$, still leaving the model response $\overrightarrow{Mod}$ virtually the same.
In such situations the minimum of the error functional $\Phi$ lies in a ``horizontal ravine'' (cf. Fig. 10 in \cite{Salamon2020}).
In the sense of Euclidean metric, a large correlation between parameters is characteristic for ill-defined optimization problems.

\begin{table}[H]
\centering
\caption{Correlation matrix for AF model with $N_{\text{branches}} = 2$.}
\begin{tabular}{c c c c c c c c}  \hline

    &$\gamma$ & $\beta$ & $c_1$& $c_2$& $\varkappa_1$& $\varkappa_2$ & $K$   \\ \hline
    $\gamma$ & 1.0000 & -0.9513 & 0.2911 & 0.0126 & -0.9438 & -0.9011 & 0.8986 \\
    $\beta$& -0.9513 & 1.0000 & -0.4874 & -0.0643 & 0.9738 & 0.9654 & -0.9644 \\
    $c_1$& 0.2911  & -0.4874 & 1.0000 & 0.2674 & -0.4783 & -0.6254 & 0.6313   \\
    $c_2$& 0.0126 & -0.0643 &  0.2674 & 1.0000 & -0.0182 & -0.0818 & 0.1130   \\
    $\varkappa_1$ & -0.9438 & 0.9738 & -0.4783 & -0.0182 & 1.0000 & 0.9840 & -0.9817   \\
    $\varkappa_2$& -0.9011 & 0.9654 & -0.6254 & -0.0818 & 0.9840 & 1.0000 & \textbf{-0.9994}   \\
    $K$ & 0.8986 & -0.9644 & 0.6313 & 0.1130 & -0.9817 & \textbf{-0.9994} & 1.0000   \\ \hline

\end{tabular} \\
\label{CorrAF2}
\end{table}

\begin{table}[H]
\centering
\caption{Correlation matrix for AF model with $N_{\text{branches}} = 3$.}
\begin{tabular}{c c c c c c c c c c}  \hline

    & $\gamma$ & $\beta$ & $c_1$ & $c_2$ & $c_3$ & $\varkappa_1$ & $\varkappa_2$ & $\varkappa_3$ & $K$ \\ \hline
    $\gamma$ & 1.0000 & -0.9513 & 0.2889 & 0.0115 & 0.0185 & -0.9431 & -0.8997 & -0.9022 & 0.8988 \\
    $\beta$ & -0.9513 & 1.0000 & -0.4847 & -0.0309 & -0.0838 & 0.9736 & 0.9647 & 0.9655 & -0.9643 \\
    $c_1$ & 0.2889 & -0.4847 & 1.0000 & 0.0889 & 0.3541 & -0.4771 & -0.6258 & -0.6178 & 0.6277 \\
    $c_2$ & 0.0115 & -0.0309 & 0.0889 & 1.0000 & 0.5741 & -0.0110 & -0.0392 & -0.0264 & 0.0546 \\
    $c_3$ & 0.0185 & -0.0838 & 0.3541 & 0.5741 & 1.0000 & -0.0295 & -0.1266 & -0.1011 & 0.1408 \\
    $\varkappa_1$ & -0.9431 & 0.9736 & -0.4771 & -0.0110 &-0.0295 & 1.0000 & 0.9833 & 0.9856 & -0.9823 \\
    $\varkappa_2$ & -0.8997 & 0.9647 & -0.6258 & -0.0392 & -0.1266 & 0.9833 & 1.0000 & \textbf{0.9997} & \textbf{-0.9997} \\
    $\varkappa_3$ & -0.9022 & 0.9655 & -0.6178 & -0.0264 & -0.1011 & 0.9856 & \textbf{0.9997} & 1.0000 & \textbf{-0.9991} \\
    $K$ & 0.8988 & -0.9643 & 0.6277 & 0.0546 & 0.1408 & -0.9823 & \textbf{-0.9997} & \textbf{-0.9991} & 1.0000 \\ \hline

\end{tabular} \\
\label{CorrAF3}
\end{table}

\begin{table}[H]
\centering
\caption{Correlation matrix for AF model with $N_{\text{branches}} = 4$.}
\begin{tabular}{c c c c c c c c c c c c}  \hline

    & $\gamma$ & $\beta$ & $c_1$ & $c_2$ & $c_3$ & $c_4$ & $\varkappa_1$ & $\varkappa_2$ & $\varkappa_3$ & $\varkappa_4$ & $K$ \\ \hline
    $\gamma$ & 1.0000 & -0.9513 & \textbf{1.0000} & 0.2856 & 0.0097 & 0.0108 & -0.0550 & -0.9429 & -0.8991 & -0.9011 & 0.8987 \\
    $\beta$ & -0.9513 & 1.0000 & -0.9514 & -0.4814 & -0.0132 & -0.0616 & 0.0488 & 0.9736 & 0.9645 & 0.9653 & -0.9644 \\
    $c_1$ & \textbf{1.0000} & -0.9514 & 1.0000 & 0.2860 & 0.0097 & 0.0108 & -0.0550 & -0.9431 & -0.8993 & -0.9013 & 0.8990 \\
    $c_2$ & 0.2856 & -0.4814 & 0.2860 & 1.0000 & 0.0139 & 0.2677 & -0.0080 & -0.4733 & -0.6238 & -0.6187 & 0.6246 \\
    $c_3$ & 0.0097 & -0.0132 & 0.0097 & 0.0139 & 1.0000 & 0.2000 & 0.0331 & -0.0086 & -0.0155 & -0.0108 & 0.0193 \\
    $c_4$ & 0.0108 & -0.0616 & 0.0108 & 0.2677 & 0.2000 & 1.0000 & -0.0067 & -0.0165 & -0.1031 & -0.0791 & 0.1096 \\
    $\varkappa_1$ & -0.0550 & 0.0488 & -0.0550 & -0.0080 & 0.0331 & -0.0067 & 1.0000 & 0.0465 & 0.0435 & 0.0435 & -0.0438 \\
    $\varkappa_2$ & -0.9429 & 0.9736 & -0.9431 & -0.4733 & -0.0086 & -0.0165 & 0.0465 & 1.0000 & 0.9827 & 0.9845 & -0.9823 \\
    $\varkappa_3$ & -0.8991 & 0.9645 & -0.8993 & -0.6238 & -0.0155 & -0.1031 & 0.0435 & 0.9827 & 1.0000 & \textbf{0.9997} & \textbf{-0.9999} \\
    $\varkappa_4$ & -0.9011 & 0.9653 & -0.9013 & -0.6187 & -0.0108 & -0.0791 & 0.0435 & 0.9845 & \textbf{0.9997} & 1.0000 & \textbf{-0.9995} \\
    $K$ & 0.8987 & -0.9644 & 0.8990 & 0.6246 & 0.0193 & 0.1096 & -0.0438 & -0.9823 & \textbf{-0.9999} & \textbf{-0.9995} & 1.0000 \\ \hline

\end{tabular} \\
\label{CorrAF4}
\end{table}

\begin{table}[H]
\centering
\caption{Correlation matrix for OW-I model with $N_{\text{branches}} = 2$.}
\begin{tabular}{c c c c c c c }  \hline

    & $\gamma$ & $\beta$ & $c_1$ & $c_2$ & $r_1$ & $K$   \\ \hline
    $\gamma$ & 1.0000 & -0.9440 & 0.0839 & \textbf{1.0000} & 0.9162 & 0.8888   \\
    $\beta$ & -0.9440 & 1.0000 & -0.2136 & -0.9440 & -0.9412 & -0.9605   \\
    $c_1$ & 0.0839 & -0.2136 & 1.0000 & 0.0839 & 0.0224 & 0.2938   \\
    $c_2$ & \textbf{1.0000} & -0.9440 & 0.0839 & 1.0000 & 0.9162 & 0.8888   \\
    $r_1$ & 0.9162 & -0.9412 & 0.0224 & 0.9162 & 1.0000 & 0.9492   \\
    $K$ & 0.8888 & -0.9605 & 0.2938 & 0.8888 & 0.9492 & 1.0000   \\     \hline

\end{tabular} \\
\label{CorrOW-I2}
\end{table}

\begin{table}[H]
\centering
\caption{Correlation matrix for OW-I model with $N_{\text{branches}} = 3$.}
\begin{tabular}{c c c c c c c c c }  \hline

    & $\gamma$ & $\beta$ & $c_1$ & $c_2$ & $c_3$ & $r_1$ & $r_2$ & $K$   \\ \hline
    $\gamma$ & 1.0000 & -0.9480 & 0.1464 & 0.0086 & \textbf{1.0000} & 0.9262 & 0.8989 & 0.8956   \\
    $\beta$ & -0.9480 & 1.0000 & -0.3040 & -0.0502 & -0.9480 & -0.9234 & -0.9620 & -0.9625   \\
    $c_1$ & 0.1464 & -0.3040 & 1.0000 & 0.0653 & 0.1464 & 0.0281 & 0.3967 & 0.4015   \\
    $c_2$ & 0.0086 & -0.0502 & 0.0653 & 1.0000 & 0.0086 & -0.0001 & 0.0034 & 0.0875   \\
    $c_3$ & \textbf{1.0000} & -0.9480 & 0.1464 & 0.0086 & 1.0000 & 0.9262 & 0.8989 & 0.8956   \\
    $r_1$ & 0.9262 & -0.9234 & 0.0281 & -0.0001 & 0.9262 & 1.0000 & 0.9209 & 0.9161   \\
    $r_2$ & 0.8989 & -0.9620 & 0.3967 & 0.0034 & 0.8989 & 0.9209 & 1.0000 & \textbf{0.9952}   \\
    $K$ & 0.8956 & -0.9625 & 0.4015 & 0.0875 & 0.8956 & 0.9161 & \textbf{0.9952} & 1.0000   \\ \hline

\end{tabular} \\
\label{CorrOW-I3}
\end{table}

\begin{table}[H]
\centering
\caption{Correlation matrix for OW-I model with $N_{\text{branches}} = 4$.}
\begin{tabular}{c c c c c c c c c c c}  \hline

    & $\gamma$ & $\beta$ & $c_1$ & $c_2$ & $c_3$ & $c_4$ & $r_1$ & $r_2$ & $r_3$ & $K$   \\ \hline
    $\gamma$ & 1.0000 & -0.9495 & 0.0027 & 0.2072 & 0.0295 & \textbf{1.0000} & 0.8984 & 0.9293 & 0.9070 & 0.8972   \\
    $\beta$ & -0.9495 & 1.0000 & -0.0241 & -0.3729 & -0.1134 & -0.9495 & -0.9634 & -0.9030 & -0.9588 & -0.9633   \\
    $c_1$ & 0.0027 & -0.0241 & 1.0000 & 0.0143 & 0.0977 & 0.0027 & 0.0006 & -0.0000 & -0.0000 & 0.0496   \\
    $c_2$ & 0.2072 & -0.3729 & 0.0143 & 1.0000 & 0.1557 & 0.2072 & 0.4822 & 0.0329 & 0.4647 & 0.4835   \\
    $c_3$ & 0.0295 & -0.1134 & 0.0977 & 0.1557 & 1.0000 & 0.0295 & 0.1659 & -0.0001 & 0.0074 & 0.1715   \\
    $c_4$ & \textbf{1.0000} & -0.9495 & 0.0027 & 0.2072 & 0.0295 & 1.0000 & 0.8984 & 0.9293 & 0.9070 & 0.8972   \\
    $r_1$ & 0.8984 & -0.9634 & 0.0006 & 0.4822 & 0.1659 & 0.8984 & 1.0000 & 0.8787 & 0.9862 & \textbf{0.9983}   \\
    $r_2$ & 0.9293 & -0.9030 & -0.0000 & 0.0329 & -0.0001 & 0.9293 & 0.8787 & 1.0000 & 0.8918 & 0.8768   \\
    $r_3$ & 0.9070 & -0.9588 & -0.0000 & 0.4647 & 0.0074 & 0.9070 & 0.9862 & 0.8918 & 1.0000 & 0.9844   \\
    $K$ & 0.8972 & -0.9633 & 0.0496 & 0.4835 & 0.1715 & 0.8972 & \textbf{0.9983} & 0.8768 & 0.9844 & 1.0000   \\   \hline

\end{tabular} \\
\label{CorrOW-I4}
\end{table}

\begin{table}[H]
\centering
\caption{Correlation matrix for OW-II model with $N_{\text{branches}} = 2$.}
\begin{tabular}{c c c c c c c c c }  \hline

    & $\gamma$ & $\beta$ & $c_1$ & $c_2$ & $r_1$ & $r_2$ & $K$ & $m$   \\ \hline
    $\gamma$ & 1.0000 &  -0.9509 & 0.0105 & 0.2270 & 0.9024 & 0.9491 & 0.8995 & 0.2970   \\
    $\beta$ &  -0.9509 & 1.0000 & -0.0605 & -0.4154 & -0.9646 & -0.9576 & -0.9638 & -0.4806   \\
    $c_1$ & 0.0105 & -0.0605 & 1.0000 & 0.1395 & 0.0548 & 0.0004 & 0.1055 & 0.3575   \\
    $c_2$ & 0.2270 & -0.4154 & 0.1395 & 1.0000 & 0.5412 & 0.2827 & 0.5451 & 0.8602   \\
    $r_1$ & 0.9024 & -0.9646 & 0.0548 & 0.5412 & 1.0000 & 0.9575 & \textbf{0.9984} & 0.6040   \\
    $r_2$ & 0.9491 & -0.9576 & 0.0004 & 0.2827 & 0.9575 & 1.0000 & 0.9539 & 0.4123   \\
    $K$ & 0.8995 & -0.9638 & 0.1055 & 0.5451 & \textbf{0.9984} & 0.9539 & 1.0000 & 0.6169   \\
    $m$ & 0.2970 & -0.4806 & 0.3575 & 0.8602 & 0.6040 & 0.4123 & 0.6169 & 1.0000   \\ \hline

\end{tabular} \\
\label{CorrOW-II2}
\end{table}

\begin{table}[H]
\centering
\caption{Correlation matrix for OW-II model with $N_{\text{branches}} = 3$.}
\begin{tabular}{c c c c c c c c c c c }  \hline

    & $\gamma$ & $\beta$ & $c_1$ & $c_2$ & $c_3$ & $r_1$ & $r_2$ & $r_3$ & $K$ & $m$   \\ \hline
    $\gamma$ & 1.0000 & -0.9512 & 0.0007 & 0.3030 & 0.0329 & 0.8994 & 0.9631 & 0.9082 & 0.8989 & 0.3561   \\
    $\beta$ & -0.9512 & 1.0000 & -0.0080 & -0.4958 & -0.1267 & -0.9645 & -0.9524 & -0.9655 & -0.9643 & -0.5452   \\
    $c_1$ & 0.0007 & -0.0080 & 1.0000 & 0.0062 & 0.0722 & 0.0084 & -0.0003 & -0.0001 & 0.0181 & 0.0679   \\
    $c_2$ & 0.3030 & -0.4958 & 0.0062 & 1.0000 & 0.2806 & 0.6377 & 0.3279 & 0.6217 & 0.6377 & 0.9015   \\
    $c_3$ & 0.0329 & -0.1267 & 0.0722 & 0.2806 & 1.0000 & 0.1930 & 0.0038 & 0.0970 & 0.1957 & 0.4912   \\
    $r_1$ & 0.8994 & -0.9645 & 0.0084 & 0.6377 & 0.1930 & 1.0000 & 0.9309 & \textbf{0.9951} & \textbf{0.9997} & 0.6922   \\
    $r_2$ & 0.9631 & -0.9524 & -0.0003 & 0.3279 & 0.0038 & 0.9309 & 1.0000 & 0.9426 & 0.9303 & 0.4203   \\
    $r_3$ & 0.9082 & -0.9655 & -0.0001 & 0.6217 & 0.0970 & \textbf{0.9951} & 0.9426 & 1.0000 & \textbf{0.9945} & 0.6562   \\
    $K$ & 0.8989 & -0.9643 & 0.0181 & 0.6377 & 0.1957 & \textbf{0.9997} & 0.9303 & \textbf{0.9945} & 1.0000 & 0.6944   \\
    $m$ & 0.3561 & -0.5452 & 0.0679 & 0.9015 & 0.4912 & 0.6922 & 0.4203 & 0.6562 & 0.6944 & 1.0000   \\ \hline

\end{tabular} \\
\label{CorrOW-II3}
\end{table}



\begin{table}[H]
\centering
\caption{Correlation matrix for OW-II model with $N_{\text{branches}} = 4$.}
\begin{tabular}{m{0.2cm} m{1.1cm} m{1.1cm} m{1.1cm} m{1.1cm} m{1.1cm} m{1.1cm} m{1.1cm} m{1.1cm} m{1.1cm} m{1.1cm} m{1.1cm} m{1.1cm}}  \hline
             & $\gamma$ & $\beta$ & $c_1$ & $c_2$ & $c_3$ & $c_4$ & $r_1$ & $r_2$ & $r_3$ & $r_4$ & $K$ & $m$   \\ \hline
    $\gamma$ & 1.0000 & -0.9512 & 0.0006 & 0.2951 & 0.0031 & 0.0302 & 0.8993 & 0.9618 & 0.8996 & 0.9076 & 0.8991 & 0.3512   \\
    $\beta$ & -0.9512 & 1.0000 & -0.0019 & -0.4877 & -0.0171 & -0.1200 & -0.9644 & -0.9528 & -0.9645 & -0.9654 & -0.9643 & -0.5396   \\
    $c_1$ & 0.0006 & -0.0019 & 1.0000 & 0.0015 & 0.1377 & 0.0126 & 0.0015 & 0.0007 & 0.0018 & 0.0009 & 0.0050 & 0.0068   \\
    $c_2$ & 0.2951 & -0.4877 & 0.0015 & 1.0000 & 0.0137 & 0.2677 & 0.6286 & 0.3223 & 0.6286 & 0.6140 & 0.6286 & 0.8970   \\
    $c_3$ & 0.0031 & -0.0171 & 0.1377 & 0.0137 & 1.0000 & 0.1511 & 0.0273 & 0.0018 & 0.0191 & 0.0019 & 0.0369 & 0.1315   \\
    $c_4$ & 0.0302 & -0.1200 & 0.0126 & 0.2677 & 0.1511 & 1.0000 & 0.1860 & 0.0031 & 0.1840 & 0.0924 & 0.1868 & 0.4777   \\
    $r_1$ & 0.8993 & -0.9644 & 0.0015 & 0.6286 & 0.0273 & 0.1860 & 1.0000 & 0.9328 & \textbf{0.9999} & \textbf{0.9952} & \textbf{0.9998} & 0.6874   \\
    $r_2$ & 0.9618 & -0.9528 & 0.0007 & 0.3223 & 0.0018 & 0.0031 & 0.9328 & 1.0000 & 0.9332 & 0.9438 & 0.9326 & 0.4206   \\
    $r_3$ & 0.8996 & -0.9645 & 0.0018 & 0.6286 & 0.0191 & 0.1840 & \textbf{0.9999} & 0.9332 & 1.0000 & \textbf{0.9955} & \textbf{0.9997} & 0.6860   \\
    $r_4$ & 0.9076 & -0.9654 & 0.0009 & 0.6140 & 0.0019 & 0.0924 & \textbf{0.9952} & 0.9438 & \textbf{0.9955} & 1.0000 & \textbf{0.9950} & 0.6525   \\
    $K$ & 0.8991 & -0.9643 & 0.0050 & 0.6286 & 0.0369 & 0.1868 & \textbf{0.9998} & 0.9326 & \textbf{0.9997} & \textbf{0.9950} & 1.0000 & 0.6882   \\
    $m$ & 0.3512 & -0.5396 & 0.0068 & 0.8970 & 0.1315 & 0.4777 & 0.6874 & 0.4206 & 0.6860 & 0.6525 & 0.6882 & 1.0000   \\ \hline

\end{tabular} \\
\label{CorrOW-II4}
\end{table}

\section*{Appendix C: Implementation of Sobol's sequence}

For each draw of the Monte Carlo method, the stochastic model \eqref{StochModel} requires 40 independent random numbers  $\sigma_k \in \mathcal{N}(0,\sigma^2)$ (20 numbers to obtain noisy data for each test). Within the quasi Monte Carlo method, they are obtained in the following way.
First, we set the properties of the Sobol sequence.
$Dimensions$ is the	number of terms of the Sobol sequence in each draw; we use $Dimensions =40$.
$Skip$ is the number of initial points to omit from Sobol's sequence, we put $Skip = 10^3$. $Leap$ is the interval between points of the sequence; $Leap = 3 \cdot 10^2$ in our case.

Recall that $N_{\text{noise}}$ is the number of draws.
Calling Sobol' generator \cite{Bratley1988} we obtain a matrix $S \in \mathbb{R}^{N_{\text{noise}} \times Dimensions}$ of quasi-random numbers uniformly distributed over the interval $[0,1]$. Then for the $j$th draw, the corresponding quasi-random variables with the normal distribution are:
\begin{equation}\label{SobolUniformNormal}
\sigma_{j,2i+1} = \cos({2\pi} S_{j,2i+1}) \cdot \sqrt{-2 \ln {S_{j,2i+1}}}, \quad
\sigma_{j,2i+2} = \sin({2\pi} S_{j,2i+1}) \cdot \sqrt{-2 \ln {S_{j,2i+1}}},
\end{equation}
for $j = 1, 2, ..., N_{\text{noise}}$, $i = 1, 2, ..., Dimensions$.

\section*{Appendix D: Fast computation of the distance}

The fast computation of the distance between two sets of parameters is based on the linearization of the strain response function $\varepsilon_{11}(t)$ with respect to the material parameters. For the fixed stress-controlled loading history (Fig. \ref{figLoadingProgramForMetric})  we evaluate the derivative
\begin{equation}
 d \mathbf{\varepsilon}/d \vec{p} (t) = \frac{\partial{\varepsilon_{1 1}(t,\vec{p})} }{\partial{\vec{p}}}|_{\vec{p} = \vec{p}^{\ \ast}}, \quad t \in [0,T_{\text{metric}}].
\end{equation}
For the parameter set $\vec{p}$ close to the center of the cloud $\vec{p}^{\ \ast}$, the axial strain $\varepsilon_{1 1}(t,\vec{p})$ is approximated as
\begin{equation}
 \varepsilon_{1 1}(t,\vec{p}) = \varepsilon_{1 1}(t,\vec{p}^{\ \ast}) +  d \mathbf{\varepsilon}/d \vec{p} (t) \cdot (\vec{p} - \vec{p}^{\ \ast}),
\end{equation}
where $\varepsilon_{1 1}(t,\vec{p}^{\ \ast})$ is the strain history related to the center of the parameter cloud. Then the mechanics-based distance between $\vec{p}^{\ \ast}$ and
 $\vec{p}$  is
\begin{equation}
 \text{dist}(\vec{p}, \vec{p}^{\ \ast}) = \max\limits_{t \in [0,T_{\text{metric}}]} | d \mathbf{\varepsilon}/d \vec{p} (t) \cdot (\vec{p} - \vec{p}^{\ \ast}) |.
\end{equation}

\bibliographystyle{plain}

\begin{thebibliography}{10}

\bibitem{AbdelKarimOhno2000}
M.~Abdel-Karim and N.~Ohno.
\newblock Kinematic hardening model suitable for ratchetting with steady-state.
\newblock {\em International Journal of Plasticity}, 16:225--240, 2000.

\bibitem{Adamus2015}
J.~Adamus and P.~Lacki.
\newblock Numerical simulation of forming titanium drawn part.
\newblock {\em Meccanica}, 51:391--400, 2016.

\bibitem{AvrilEtAl2004}
S.~Avril, M.~Gr\'ediac, and F.~Pierron.
\newblock Sensitivity of the virtual field method to noisy data.
\newblock {\em Computational Mechanics}, 34:439 -- 452, 2004.

\bibitem{Bartel2016}
T.~Bartel, M.~Osman, and A.~Menzel.
\newblock A phenomenological model for the simulation of functional fatigue in
  shape memory alloy wires.
\newblock {\em Meccanica}, 52(4-5):973 -- 988, 2017.

\bibitem{Beck2007}
J.V. Beck and K.J. Arnold.
\newblock {\em Parameter Estimation in Engineering and Science}.
\newblock John Wiley and Sons, 2007.

\bibitem{Benedetti2020}
K.C.B. Benedetti, P.B. Gonçalves, and F.M.A. Silva.
\newblock Nonlinear oscillations and bifurcations of a multistable truss and
  dynamic integrity assessment via a monte carlo approach.
\newblock {\em Meccanica}, 55:2623 -- 2657, 2020.

\bibitem{Benedix2000}
U.~Benedix.
\newblock {\em Parametrschätzung für elastisch-plastische Deformatiosgesetze
  bei Berücksichtigung lokaler und globaler Vergleichsgrößen. Dissertation}.
\newblock Chemnitz University, 2000.

\bibitem{Bratley1988}
P.~Bratley and B.L. Fox.
\newblock Algorithm 659: Implementing sobol's quasirandom sequence generator.
\newblock {\em ACM Transactions on Mathematical Software}, 14(1):88--100, 1988.

\bibitem{Bruenig2008}
M.~Brünig, O.~Chyra, D.~Albrecht, L.~Driemeier, and M~Alves.
\newblock A ductile damage criterion at various stress triaxialities.
\newblock {\em Int. J. Plasticity}, 24(10):1731--1755, 2008.

\bibitem{Collins1993}
J.A. Collins.
\newblock {\em Failure of materials in mechanical design: analysis, prediction,
  prevention}.
\newblock John Wiley \& Sons, 1993.

\bibitem{Francois2001}
M.~François.
\newblock A plasticity model with yield surface distortion for non proportional
  loading.
\newblock {\em Int. J. Plasticity}, 17:703--717, 2001.

\bibitem{Grediac2004}
M.~Gr\'ediac and F.~Pierron.
\newblock Applying the virtual fields method to the identification of
  elasto-plastic constitutive parameters.
\newblock {\em International Journal of Plasticity}, 22:602--627, 2004.

\bibitem{Harth2007}
T.~Harth and J.~Lehn.
\newblock Identification of material parameters for inelastic constitutive
  models using stochastic methods.
\newblock {\em GAMM-Mitt.}, 30(2):409--429, 2007.

\bibitem{Harth2004}
T.~Harth, S.~Schwan, J.~Lehn, and F.G. Kollmann.
\newblock Identification of material parameters for inelastic constitutive
  models: statistical analysis and design of experiments.
\newblock {\em International Journal of Plasticity}, 20:1403--1440, 2004.

\bibitem{Hashiguchi1989}
K.~Hashiguchi.
\newblock Subloading surface model in unconventional plasticity.
\newblock {\em International journal of solids and structures}, 25(8):917--945,
  1989.

\bibitem{Hashiguchi2017}
K.~Hashiguchi.
\newblock {\em Foundations of elastoplasticity: subloading surface model}.
\newblock 2017.

\bibitem{Haupt2013}
P.~Haupt.
\newblock {\em Continuum mechanics and theory of materials}.
\newblock Springer Science \& Business Media, 2013.

\bibitem{Kang2009}
G.~Kang, Y.~Liu, J.~Ding, and Q.~Gao.
\newblock Uniaxial ratcheting and fatigue failure of tempered 42crmo steel:
  Damage evolution and damage-coupled visco-plastic constitutive model.
\newblock {\em International Journal of Plasticity}, 25(5):838--860, 2009.

\bibitem{Kaygorodtseva2020}
A.A. Kaygorodtseva, V.I. Kapustin, K.V. Zakharchenko, and A.V. Shutov.
\newblock On the ratcheting of the vt6 alloy in a range of loading scenarios.
\newblock {\em Journal of Physics: Conference Series}, 1666 (2020) 012020,
  2020.

\bibitem{Lemaintre1984}
J.~Lemaitre.
\newblock A three-dimensional ductile damage model applied to deep-drawing
  forming limits.
\newblock {\em Mechanical Behaviour of Materials}, pages 1047--1053, 1984.

\bibitem{Lion2000}
A.~Lion.
\newblock Constitutive modelling in finite thermoviscoplasticity: a physical
  approach based on nonlinear rheological elements.
\newblock {\em International Journal of Plasticity}, 16:469--494, 2000.

\bibitem{LevenbergMarquardt2005}
M.I.A. Lourakis.
\newblock A brief description of the {L}evenberg-{M}arquardt algorithm
  implemented by levmar.
\newblock {\em Foundation of Research and Technology}, 4(1):1, 2005.

\bibitem{NelderMead1965}
J.A. Nelder and R.~Mead.
\newblock A simplex method for function minimization.
\newblock {\em The computer journal}, 7(4):308--313, 1965.

\bibitem{Niederreiter1978}
H.~Niederreiter.
\newblock Quasi-monte carlo methods and pseudo-random numbers.
\newblock {\em Bulletin of the American mathematical society}, 84(6):957--1041,
  1978.

\bibitem{OhnoWang1993}
N.~Ohno and J.D. Wang.
\newblock Kinematic hardening rules with critical state of dynamic recovery,
  part i: formulation and basic features for ratchetting behavior.
\newblock {\em International journal of plasticity}, 9(3):375--390, 1993.

\bibitem{Salamon2020}
R.~Salamon, H.~Kami\'nski, and P.~Fritzkowski.
\newblock Estimation of parameters of various damping models in planar motion
  of a pendulum.
\newblock {\em Meccanica}, 2020.

\bibitem{Seibert2000}
T.~Seibert, J.~Lehn, S.~Schwan, and F.G. Collmann.
\newblock Identification of material parameters for inelastic constitutive
  models: Stochastic simulations for the analysis of deviations.
\newblock {\em Continuum Mech. Thermodyn}, 12:95--120, 2000.

\bibitem{ShutovIhlemann2011}
A.~V. Shutov and J.~Ihlemann.
\newblock On the simulation of plastic forming under consideration of thermal
  effects.
\newblock {\em Materialwissenschaft und Werkstofftecnhik}, 42 (7):632--638,
  2011.

\bibitem{ShutovIhlemann2012}
A.~V. Shutov and J.~Ihlemann.
\newblock A viscoplasticity model with an enhanced control of the yield surface
  distortion.
\newblock {\em International Journal of Plasticity}, 39:152--167, 2012.

\bibitem{ShutovSilbermannInhelamm2015}
A.~V. Shutov, C.~B. Silbermann, and J.~Ihlemann.
\newblock Ductile damage model for metal forming simulations including refined
  description of void nucleation.
\newblock {\em International Journal of Plasticity}, 71:195--217, 2015.

\bibitem{ShutovKaygorodtsevaZAMM2019}
A.V. Shutov and A.A. Kaygorodtseva.
\newblock Parameter identification in elasto‐plasticity: distance between
  parameters and impact of measurement errors.
\newblock {\em ZAMM Journal of Applied Mathematics and Mechanics}, 99(8), 2019.

\bibitem{ShutovKaygorodtsevaActaMech2020}
A.V. Shutov and A.A. Kaygorodtseva.
\newblock Sample shapes for reliable parameter identification in
  elasto-plasticity.
\newblock {\em Acta Mech}, 2020.

\bibitem{ShutovKreissig2008}
A.V. Shutov and R.~Krei{\ss}ig.
\newblock Finite strain viscoplasticity with nonlinear kinematic hardening:
  Phenomenological modeling and time integration.
\newblock {\em Computer Methods in Applied Mechanics Engineering},
  197:2015--2029, 2008.

\bibitem{ShutovLarichkin2017}
A.V. Shutov, A.Y. Larichkin, and V.A. Shutov.
\newblock Modelling of cyclic creep in the finite strain range using a nested
  split of the deformation gradient.
\newblock {\em ZAMM‐Journal of Applied Mathematics and Mechanics/Zeitschrift
  für Angewandte Mathematik und Mechanik}, 97(9):1083--1099, 2017.

\bibitem{ShutovPanhans2011}
A.V. Shutov, S.~Panhans, and R.~Krei{\ss}ig.
\newblock A phenomenological model of finite strain viscoplasticity with
  distortional hardening.
\newblock {\em ZAMM}, 91(8):653--680, 2011.

\bibitem{Sobol1967}
I.M. Sobol.
\newblock Distribution of points in a cube and approximate evaluation of
  integrals.
\newblock {\em Comput. Maths. Math. Phys.}, 7:86--112, 1967.

\bibitem{Soize2017}
C.~Soize.
\newblock {\em Uncertainty Quantification}.
\newblock Springer, 2017.

\bibitem{Surmiri2019}
A.~Surmiri, A.~Nayebi, and H.~Rokhgireh.
\newblock Application of anisotropic continuum damage mechanics in ratcheting
  characterization.
\newblock {\em Mechanics of Advanced Materials and Structures}, 2020.

\bibitem{Vladimirov2008}
I.N. Vladimirov, M.P. Pietryga, and S.~Reese.
\newblock On the modelling of non‐linear kinematic hardening at finite
  strains with application to springback—comparison of time integration
  algorithms.
\newblock {\em International Journal for Numerical Methods in Engineering},
  75(1):1--28, 2008.

\bibitem{Wolszczak2019}
P.~Wolszczak, P.~Lonkwic, A.~Cunha~Jr., L.~Litak, and S.~Molski.
\newblock Robust optimization and uncertainty quantification in the nonlinear
  mechanics of an elevator brake system.
\newblock {\em Meccanica}, 54:1057--1069, 2019.

\bibitem{WrightNocedal1999}
S.~Wright and J.~Nocedal.
\newblock {\em Numerical optimization}.
\newblock Springer Science, 1999.

\bibitem{Yang2005}
X.~Yang.
\newblock Low cycle fatigue and cyclic stress ratcheting failure behavior of
  carbon steel 45 under uniaxial cyclic loading.
\newblock {\em International Journal of Fatigue}, 27(9):1124--1132, 2005.

\bibitem{Zhu2017}
S.~P. Zhu, Q.~Lei, and Q.Y. Wang.
\newblock Mean stress and ratcheting corrections in fatigue life prediction of
  metals.
\newblock {\em Fatigue \& Fracture of Engineering Materials \& Structures},
  40(9):1343--1354, 2017.

\end{thebibliography}

\end{document}